\documentclass[preprintnumbers,eqsecnum,nofootinbib,floatfix]{revtex4}
\usepackage{amsmath,amsthm,amssymb}
\usepackage[dvips]{graphics,lscape,rotating}
\usepackage{psfrag}

\newcommand\be{\begin{displaymath}}
\newcommand\ee{\end{displaymath}}
\newcommand\beq{\begin{equation}}
\newcommand\eeq{\end{equation}}
\newcommand\bea{\begin{eqnarray*}}
\newcommand\eea {\end{eqnarray*}}
\newcommand\intersect{\cap}
\newcommand\past{\mathrm{past}}
\newcommand\fut{\mathrm{fut}}
\newcommand\vol{\mathrm{Vol}}
\def\la{{\langle}}
\def\ra{{\rangle}}
\def\a{{\alpha}}

\newcommand\ltsim{\stackrel{<}{\scriptstyle\sim}}

\def\f{{\mathbf f}}
\def\x{{\mathbf x}}
\def\y{\mathbf{y}}
\def\p{\mathbf{p}}
\newcommand{\causet}{\mathcal{P}}
\newcommand{\eg}{\emph{e.g.}\,}
\newcommand{\ie}{\emph{i.e.}\,}
\newcommand{\cf}{\emph{cf.}\,}

\newcommand{\spacelikepoints}{spacelike points}
\newcommand{\bgdist}{naive spatial distance}
\newcommand{\Bgdist}{Naive spatial distance}
\newcommand{\lgdist}{generalized spatial distance}
\newcommand{\whp}{almost surely}
\newcommand{\meff}[1]{m_#1^{\mathrm{eff}}}
\newcommand{\nlinks}[1]{\langle \# \; #1-\textrm{links} \rangle}
\newcommand{\M}{\mathbb{M}}

\newcommand{\rf}{ref.\ } 
\newcommand{\threshold}{\lambda} 
 
\newcommand{\slink}{s-link} 

\newcommand{\spover}{spacelike overestimation}
\newcommand{\tiunder}{timelike underestimation}
\newcommand{\voldist}{volume distance}
\newcommand{\Voldist}{Volume distance}
\newcommand{\minpair}{minimizing pair}
\newcommand{\contminpair}{continuum minimizing pair}

\newcommand{\cqg}{\emph{Class.\ Quant.\ Grav.}}
\newcommand{\eprint}[1]{$\langle$e-print arXiv: #1$\rangle$}

\begin{document}

\title{Spacelike distance from discrete causal order}

\author{David Rideout}

\affiliation{Perimeter Institute for Theoretical Physics \\ 31
Caroline Street North, Waterloo, Ontario N2L 2Y5, Canada}

\author{Petros Wallden}

\affiliation{ Raman Research Institute \\
 Sadashivanagar, Bangalore - 560 080, India}

\date{\today}

\begin{abstract}

Any discrete approach to quantum gravity must provide some prescription as to
how to deduce continuum properties from the discrete substructure.  In the
causal set approach it is straightforward to deduce timelike distances, but
surprisingly difficult to extract spacelike distances, because of the unique
combination of discreteness with local Lorentz invariance in that approach.
We propose a number of methods to overcome this difficulty, one of which
reproduces the spatial distance between two points in a finite region of
Minkowski space.
We provide numerical evidence that this
  definition can be used to define a `spatial nearest neighbor' relation on a
  causal set, and conjecture that 
this can be exploited to define the length
  of `continuous curves' in causal sets which are approximated by curved
  spacetime.
  This provides some evidence in support of the
``Hauptvermutung'' of causal sets.

\end{abstract}

\maketitle
\tableofcontents

\section{Introduction}

One of the central problems which confronts every discrete/combinatorial
approach to quantum gravity (see for example \cite{myrheim, thooft, blms,
  cdts, lqg, spin_foams, graphity, rbw})
is that of the so-called `inverse
problem' \cite{inverse_problem}, which may be decomposed into two aspects.
One regards the question as to how smooth continuum-like structures emerge
from the underlying discrete dynamics.  Implicit in this is a second,
that of how to 
recognize that we have
something which is smooth and continuum-like, if we did?  The paper
addresses this second question,
by proposing a means
to extract spatial distances and other geometric information from a
causal set.  Along with earlier prescriptions for measuring timelike
distances from a discrete causal order \cite{myrheim, bg}\footnote{This list of references is by no means complete.  Timelike distance
  in a causal set has been studied from wide variety of perspectives,
  including the study of random permutations, the mathematics of random order
  structures, the optimization of algoirthms for accessing data off hard
  drives, and efficiency of airplane boarding strategies.  For a more complete
  survey of the vast literature on this subject, see \cite{bachmat} and the
  references therein.}, these results give promise to specifying the full spacetime geometry (aka
metric) from a
causal set. This contributes 
 towards the
`Hauptvermutung' (central conjecture) of causal set theory, which
claims that two distinct, non-isometric spacetimes cannot arise from
a single causal set.

\subsection{Causal Sets}

Causal Sets is one of the major approaches to constructing a theory
of quantum gravity, and is perhaps the most minimalist in its
starting point. The causal set hypothesis asserts that a discrete
set of `atoms of spacetime', endowed with merely a partial order relation, underlies
the continuum, and provides the basis for a consistent quantum
description of spacetime.
The basic assumptions are that spacetime is
\emph{fundamentally discrete}, and that to this one need add only
causal ordering of the elements to recover the full spacetime structure,
including dimension, topology, differential structure, and metric, at
macroscopic scales.

Several of the recent developments that have occurred in this approach
at the kinematical level are
\cite{lorentz_invariance, homology, faithful_embedding,reid_geodesics}, at dynamical level
\cite{Brightwell:2007,johnston}, and at the
phenomenological level \cite{Sorkin:2007,Sorkin:dalembertian}.
For reviews 
see \eg \cite{discrete_lambda, causet_reviews, valdivia, joe_review}. Perhaps the greatest
piece of evidence in support of the hypotheses underlying the causal set program
to date
is the successful
prediction of the order of magnitude of the cosmological constant. In 1990
Rafael Sorkin argued that,
from very general principles of spacetime discreteness, and the
complementarity of spacetime volume and a cosmological constant in the
gravitational action, one should expect a non-zero, fluctuating cosmological
constant, whose magnitude was at the time just beyond the bounds of experimental
observation \cite{discrete_lambda, Ahmed:2002} (and believed to be vanishing by 
much of the scientific community).
Note that this implied that an exactly vanishing cosmological
constant was highly unlikely (if not ruled out) by causal sets.

Mathematically a causal set (or causet) is a set $C$ endowed with a partial
order relation $\prec$ which is irreflexive ($x \nprec x$),
transitive ($x \prec y \prec z \implies x \prec z$), and locally
finite $[x,z] \equiv \left(|\{y | x \prec y \prec z \}| < \infty \; \forall\quad
x, z \in C\right)$ (where $|A|$ indicates cardinality of the set $A$).  The local finiteness condition imposes
the requirement of discreteness, because it requires that there is only a
finite number of elements between every pair in the causet.
The quantity $[x,y]$ used to define local finiteness, which is the collection
of elements following $x$ but preceeding $y$ in the partial order, is called a
\emph{causal interval} or simply \emph{interval}, and plays an important role
in the following.  It is also sometimes called an Alexandrov neighborhood or
Alexandrov set.
A \emph{link} is a relation between two
elements which is not implied by transitivity, an `irreducible
relation'. A \emph{chain} is a sub-causal set for which every pair
of elements is related.  An \emph{antichain} is a subcauset which
contains no relations.  The term \emph{$n$-antichain} is shorthand for an
antichain containing $n$ elements.  The past of an element $x$ $\past(x)$ is
the set of elements $y\in C$ such that $y \prec x$.  Similarly the future of
an element $x$ $\fut(x)$ is
the set of elements $y\in C$ such that $x \prec y$.
An element is called \emph{minimal} if its past is empty.
An element is called \emph{maximal} if its future is empty.

Causal sets are fundamentally discrete, which means that one does not expect
the spacetime continuum to arise as a
continuum limit of the theory.  What we should be
seeking is a continuum \emph{approximation}, which is some
continuum spacetime that approximates the underlying fundamental discrete
structure. 
To this end, we say that a
particular spacetime approximates a causal set via the definition of a
\emph{faithful embedding}.  A faithful embedding is a map $\phi$ from a causal set
$C$ to a spacetime $M$ 
that \emph{(a)} preserves the
causal relation (\ie{} $x\prec y\iff \phi(x)\prec \phi(y)$) and
\emph{(b)} is `volume preserving', meaning that the number of elements mapped
to every spacetime region
is Poisson distributed, with mean the volume of
the spacetime region in fundamental units,
and \emph{(c)} $M$ does not possess curvature at scales smaller than that
defined by the `intermolecular spacing' of the embedding.
As mentioned above, the central conjecture of
causal set theory is that a causal set cannot be \emph{faithfully}
embedded into two distinct, non-isometric spacetimes.
However, the
recovery of the corresponding metric and of some
analogue of the differential structure from the causal set is still an open problem.

Throughout the paper we will need to refer to elements both of the embedding
manifold and of the embedded causal set.  To help clarify the discussion, 
we will reserve the term `points' for the events of spacetime, and use
`element' to refer to the elements of the causal set.

The results presented here are purely kinematical in nature, that is they
make no reference to any particular scheme for formulating a quantum dynamics
for causal sets.  Thus we expect them to be useful regardless of one's
attitude toward `quantization'.
The notion of a faithful embedding provides a relatively concrete
prescription
with which to associate a continuum spacetime with a causal set, at the kinematical level.  This
concreteness is useful in
allowing one to address kinematical questions, such as recovery of spatial
distance from a causal set,
independently of
one's approach toward formulating quantum dynamics.
In particular, since local Lorentz invariance already holds at the microscopic
level,
there is no need to recover it dynamically via some `quantum superposition'.
In fact it has been argued that \emph{any} fundamentally discrete approach to
quantum gravity must in some way 
contain a combinatorial structure which closely resembles a causal set, if
that theory hopes to describe a continuum which posseses the macroscopic
symmetries of Minkowski space \cite{requires_causets}.

Most of the results described in this paper are for causal sets which are
well approximated by Minkowski spacetime.
We point out that Minkowski spacetime is
very important on its own right, for a variety of reasons:
\begin{itemize}
\item Curved spacetime is locally flat. Therefore the results obtained are
  important for understanding the small scales of causal sets that are
  approximated by a curved 
spacetime.\footnote{Note that
    this `small scale' may be after some coarse graining of the causal set, to
    get to the macroscopic, continuum-like scales.  It is of course quite open
  what the smallest scales of the causal set will look like.}
\item Most of particle physics is done for quantum matter in flat spacetime.
Thus our first attempts at defining quantum field theory on a causal set
background will be on causal sets which faithfully embed into
Minkowski space. A definition of spatial distance on a causal set
can be useful in formulating the dynamics, \eg in writing down an
action for fields on a causet background \cite{sverdlov}.  An alternative
approach is to write down an action based upon non-local but causal
differential operators
defined on the causal set, as described in \cite{Sorkin:dalembertian, joe_review}.
Additionally, recent progress has been made in formulating the causal 
propagator for massive particles propagating on a causal set \cite{johnston}.
Perhaps a better understanding of spatial distance can be of use in
formulating the Feynmann propagator.

\item 
For the universe in which we live, spacetime is almost flat at the largest
scales.

\item In a model of full 2-$d$ quantum gravity of causal sets
  \cite{Brightwell:2007}, the main contribution from the sum over histories
  comes from causets which correspond to an interval in $\M^2$.
  While this is a toy model, 
  it provides another reason to be interested in causal sets which are
  approximated by Minkowski spacetime.
\end{itemize}

\subsection{Outline of this paper}

In section \ref{Section 2} we review 
the well understood notion of timelike distance on a causal set, a first attempt at defining spatial
distance, and the reason for the latter's failure. In section \ref{Section 3} we explore numerically the
aforementioned timelike and spacelike distance measures. In section \ref{Section 4} we proceed by considering
some spatial geometrical information that we \emph{can} recover from a causal
set. We define
$n$-links and discuss the dimension and `manifoldness' information they may contain.
We define `\lgdist', 
which is the radius of the minimum bounding sphere of $d$ points
in
$d$-dimensions. Using a definition of `equidistant elements',
we recover the `sphere distance', which is the diameter of a sphere
circumscribed by $d$ points
in $d$-spacetime dimensions. All these definitions are tested
numerically. 
Then in 
section \ref{Section 5} 
we
suggest a new proposal for spatial distance, which 
does not
suffer from the problems of previous suggestions.
This is the main result of the paper. In
section \ref{Section 6} we show how we can compute closest neighbors from this
distance. 
From this definition we may be able to define the length of continuous (timelike or
spacelike) curves on a causal set, which generalizes to curved spacetime.
We summarize and conclude in
section \ref{Section 7}.

\section{Elaborations on earlier ideas}\label{Section 2}
Given a spacetime manifold, it is easy to generate a causal set which will
faithfully embed into it.  One simply `sprinkles' elements uniformly
(with respect to the volume measure) at random, and then deduces the
relations among the elements from the causal structure of the spacetime.  Going
the other way is much more difficult.  One of the major tasks of causal set
kinematics is to learn to deduce properties of the approximating continuum
merely from the order relation.

The recovery of the 
spacetime manifold that approximates the
causal set, using only the partial ordering of the elements,
has not
been fully achieved.  Much is understood about how to compute the dimension
\cite{dimension, valdivia},
and more recently the spatial topology \cite{homology}. In the following, we will review some early works
on the timelike distance (proper time) \cite{thooft,myrheim, bg} and
spatial distance, since this is what we will be dealing with in the
rest of the paper.

\subsection{Timelike distance}
\label{timelike_dist_theory}
Following \cite{bg}, we define proper time $d(x,y)$, between two
related elements $x\prec y$, to be the number of links $L$ in the
\emph{longest}
 chain between (and including) $x$ and $y$\footnote{Note that the Lorentzian character of the partial order is manifested in this definition.
Graphs (of `finite valence'), on the other hand, naturally embed into Euclidean spaces, and hence one generally defines distance in terms of \emph{shortest} paths on a graph.}.
\be d(x,y):=L\ee
Thus to compute this, 
one considers all chains starting from $x$ and ending at $y$, and 
counts the number of links in (one of\footnote{For a sprinkling into Minkowski
  space, the number of longest chains will generally be very large, in fact it
  grows exponentially with the length of the chain.})
the largest one(s).
This definition
is intrinsic to the causal set, and does not depend on whether it
can be faithfully embedded into a manifold, nor on the expected dimension of
such a
manifold. In \cite{bg} (and references therein) it is shown that, in the case of a
casual set $C$ which
arises by a sprinkling of density $\rho$ into $d$-dimensional Minkowski
space\footnote{Note that we are abusing notation here, by using the same
  symbol $d$ for the dimension of Minkowski space and distance.  We expect
  that the meaning will be clear from context.}
(so one gets a faithful embedding $\phi: C \to M$), the
distance $d(x,y)$ is proportional to the proper time between the
endpoints $\phi(x)$ and $\phi(y)$.  In particular
the authors state that
\beq
L(\rho V)^{-1/d}\rightarrow m_d \;\;\; \mathrm{ as } \;\;\; \rho V\rightarrow \infty \;,
\label{bg_result}
\eeq
for some constant $m_d$ which depends upon the dimension. Here $V$
is the spacetime volume of the causal interval $J^+(x) \cap J^-(y)$
($J^\pm(x)$ represents the causal future/past of $x$ respectively).
The
exact value of $m_2$ is known to be $2$, while for other dimensions
some bounds exist: $1.77\leq m_d\leq 2.62$. 

It is known that in two dimensions the variance in $L$ grows as $N^{1/3}$
(slower than the naive $L \sim N^{1/2}$ that one might expect) \cite{bdj}.
Furthermore, the next-to-leading order term in the asymptotic behavior
(\ref{bg_result}) is also known.  It turns out that
the expected value $\langle L \rangle$ underestimates the limit (\ref{bg_result}) by a term proportional to
$N^{-1/3}$ 
\cite{bdj}.  We make use of this fact
in fitting our numerical results of section \ref{timelike_numerics}.

~\\
Given a causet that is faithfully embedded in
$d$-dimensional Minkowski spacetime, one can easily extract
timelike distance in another manner, taking direct advantage of the
number-volume correspondence of the causal set \cite{thooft}.
Consider again a timelike separated pair of
causet elements
$x \prec y$.  Associated with this pair is an Alexandrov interval in
$\M^d$, $J^+(\phi(x))\cap J^-(\phi(y))$, whose volume $V_{xy}$ is
simply
\beq
V_{xy}:=\int_{J^+(x)\cap J^-(y)} d^dz =\eta(d)\times l^d_{xy}
\label{volume-interval}
\eeq
where $l_{xy}$ indicates the `spacetime interval', \ie the proper time
separation between $x$ and $y$.
The coefficient is given by
\be
\eta(d)=\frac{2V^s_{d-1} }{2^dd} \;,
\ee
where $V^s_n$ is the
volume of a 
unit $n$-ball. We can use this simple fact to 
compute the
proper time between causal set elements $x,y$, by counting the number of
elements in the causal interval $[x,y]$,
and inverting 
(\ref{volume-interval}). We will call this distance \emph{\voldist}.

\Voldist{} has a number of advantages over counting the length of
the longest chain, in that it is exact for finite regions, and,
since it effectively integrates over a large region of the causet,
has considerably smaller fluctuations.  Some disadvantages are that it is more
explicitly dependent upon dimension, and it is
not clear how to make use of such a definition in the context of an
inhomogeneous spacetime. We will at times take advantage of
\voldist{} in the sequel, for convenience, though for the most part
we stick with the arguably more fundamental method of counting
chains to measure timelike distance in the causal set.  In principle
either definition of timelike distance can be used in the
constructions described herein.
Note that \voldist{} is used in the construction of ref.\
\cite{faithful_embedding}, which seeks to recover the faithful embedding in $\M^d$
from the causal set, since there the accuracy for finite regions is
quite important.

\subsection{Spacelike distance}
\label{bgfailure_description}
The situation is quite different for spacelike distance.  Previously
(\eg \cite{thooft,bg}) a suggestion for what might be spacelike
distance was brought forward.  Along with this suggestion came the reason
why in general this would fail for a causal set. We will thus use
the term \emph{\bgdist{}} for this proposal, since the authors of
\cite{bg} themselves rejected it.
An important note is that we restrict our attention here to Minkowski
spacetime. In section \ref{Section 6} we will mention possible
generalizations to curved spacetime.\footnote{It should be
highlighted here that even the concept of spacelike distance itself is not
unambiguously defined for curved spacetime.  However,
the length of a spacelike curve \emph{is} well defined.}

Given two unrelated elements $x,y$ one 
defines the \bgdist{} $d_{ns}(x,y)$ to be
given by the minimum timelike distance between an element $w$ in their common past
and another $z$ in their common future. Thus 
$w\prec (x$, $y) \prec z$, and 
\be
d_{ns}(x,y):=\min_{w,z} d(w,z) \;.
\ee
The timelike distance can be
calculated using either the length of the longest chain or \voldist. The
motivation for this definition comes from the fact that in
Minkowski spacetime, the minimum timelike distance from
the common past to the common future of two spacelike points $x,y$, is
indeed their spacelike distance. This proposal works for a causal
set that is approximated by 2-$d$ Minkowski spacetime. However, it
fails in higher dimensions.

In 1+1 dimensions, for a given pair of points $x,y$, there is a unique pair of
points $w\prec x, y$ and $z \succ x, y$ (in the continuum) which are separated
by the minimum proper time.  We therefore expect to find $O(1)$ pairs of
elements $w,z$ (in the causal set) which have a minimum timelike separation,
and the smallest of these should be a good estimate for the spatial separation
between $\phi(x)$ and $\phi(y)$, for a faithful embedding $\phi:C\to\M^2$.
However, in higher dimension, there will be an infinite number of such pairs
of points $w, z$, each separated by the same minimum proper time.  It is this
infinity of pairs which causes difficulty in a discrete setting.

To see in detail why \bgdist{} fails to recover spacelike distance for causal sets
which faithfully embed into $\M^d$ for $d>2$, let us first consider the
situation in the continuum.  We have a pair of spacelike points $x, y$,
and
seek pairs of elements $w$ in their common past and $z$ in their common
future, for which the timelike distance along the geodesic connecting $w$ and
$z$ is shortest.  Such a point $w$
will necessarily lie on the intersection of $x$ and $y$'s past light cones, and
$z$ on the intersection of their future light cones.  For any such point $w$,
there will be a unique $z$ which minimizes the timelike distance.  We call
such a pair of points a \emph{\contminpair}.
As mentioned above, in
1+1 dimensions there is 
a unique such pair $w$, $z$, while in higher dimension the intersections of
the light cones will form a surface of co-dimension 2, and since for each of these points in the past
there exists a corresponding one in the future which forms a \contminpair,
we find an infinite number of such pairs.

For a causal set which is faithfully embedded into $\M^d$, it is clear that
there will be in general pairs of causal set elements `close'\footnote{The precise
  definition 
  of `close' is not important at this stage.  We can
  use a Euclidean metric on the embedding space, for example.}
to each \contminpair{}.
In section \ref{Section 5}
we will properly define a discrete analogue of a minimizing pair 
in a causal set.  For the moment we will rely on the above intuitive
description of `discrete minimizing pairs', as elements $w \in \past(x)
\intersect \past(y)$ and $z \in \fut(x) \intersect \fut(y)$ which are close to
some continuum minimizing pair.

\begin{figure}[hbtp]
  \includegraphics[width=7in]{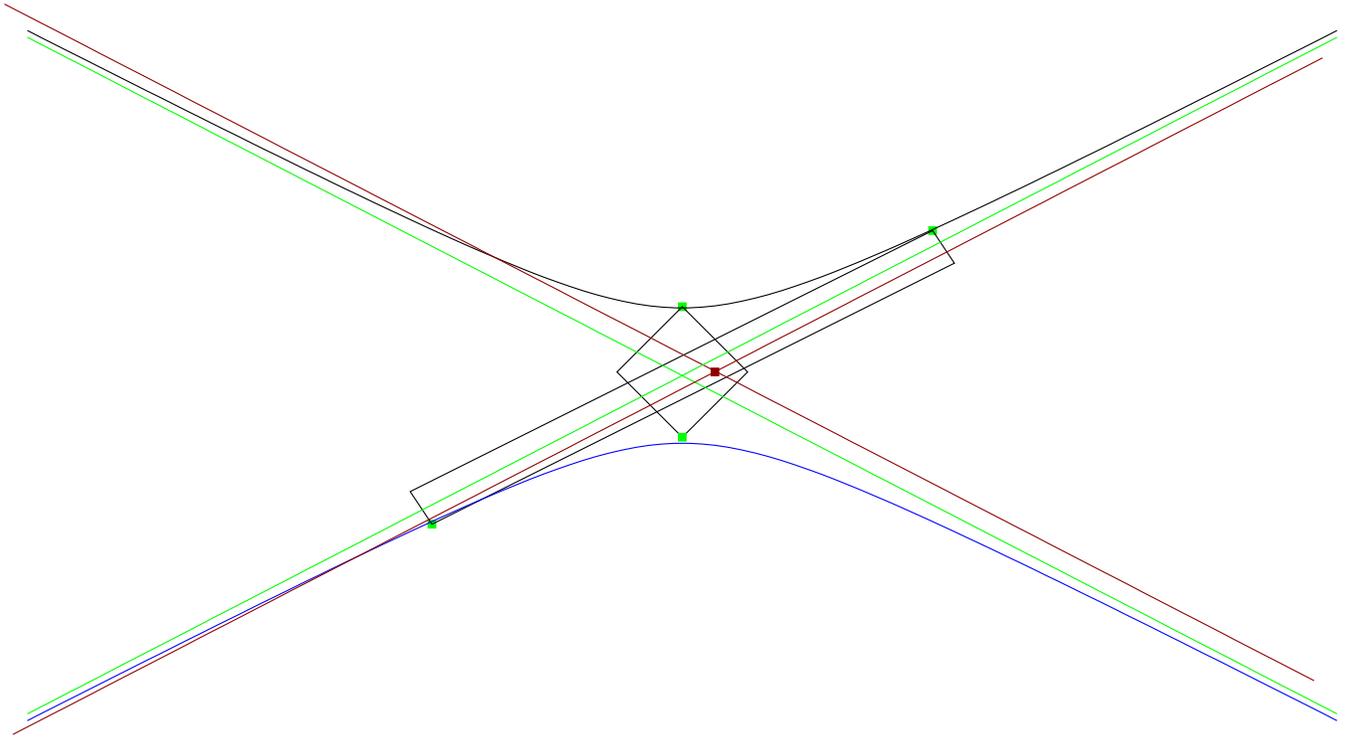}
  \caption{Illustration of two intervals associated with two minimizing pairs
    in a sprinkling into $\M^3$.
    The elements $x$ and $y$ lie at the center of the figure, but
    displaced out of and into the page by some finite equal amount.  The
    hyperbolae indicate the intersection of the past and future light cones of
    $x$ and $y$.  The long straight green lines are the asymptotes of the
    hyperbolae.  The green dots with black intervals indicate the projection
    of two causal intervals for minimizing pairs onto the plane of the page.
    Please ignore the red dot in the center of both intervals, along with the
    pair of red lines which pass through this dot.  These will be used in
    section \ref{diameters_sec}.}
  \label{minimizing_pair}
\end{figure}

An important point is that proper time for each minimizing pair $w,z$ (measured
in whichever way) depends upon
the contents of the causal interval $[w,z]$.  Since every such interval must
necessarily contain the elements $x$ and $y$, they all overlap.  However, the
extent of overlap (as measured by cardinality, or spacetime volume in the
faithful embedding) can be made arbitrarily small by considering causal
intervals which are related by a sufficiently large boost.
We call such minimizing pairs \emph{independent minimizing pairs}.  (See figure
\ref{minimizing_pair} for an illustration.)
Since this is true of any pair of such intervals $[w_1,z_1]$ and $[w_2,z_2]$,
it is clear that the infinite collection of intervals $[w_i,z_i]$ sample
an infinite volume of Minkowski space\footnote{
  In case the reader is still not convinced, let us
  be more explicit as follows:  Consider one continuum minizing
  pair $w_i, z_i$ (a pair of points $w_i \in J^-(x)\cap J^-(y)$ and $z_i \in
  J^+(x)\cap J_+(y)$ 
  whose (timelike) geodesic
  distance is minimum), with its corresponding interval $[w_i,z_i]\equiv
  J^+(w_i)\cap J^-(z_i)$, and a
  second pair $w_j, z_j$ with its interval $[w_j,z_j]$.  These intervals will
  have a finite overlap $[w_i,z_i] \intersect [w_j,z_j]$.  Now select any
  finite overlap threshold $T$. 
  There will exist an infinite number of
  pairs/intervals $[w_k,z_k]$ which have an overlap $\vol([w_i,z_i] \intersect
  [w_k,z_k]) \leq T$.  Now select one of these $[w_k,z_k]$, and repeat the
  argument, to find another interval with arbitrarily small overlap with the
  first two.  Repeating this argument an infinite number of times leads to an
  infinite number of intervals with arbitrarily small overlap among
  themselves.  (Note that one has to be careful not to have repeats in the
  sequence of intervals which this algorithm yields.  It is clear that this is
  possible, \eg in $\M^3$ by `always going in the same boost direction' to find
  new intervals.  In higher dimension there will obviously be more `arbitrarily
  independent intervals' than in $\M^3$.)}.
Thus, because the probability of a sprinkling exactly two elements into
an interval of any size is non-zero, with probability 1 we will find
some intervals (in fact an infinite number) which contain only the elements $x,
y$, and thus when minimizing over all pairs we will get a trivial distance of
2 (for either method of computing timelike distance)
for \emph{any} unrelated pair $x, y$.

Many similar proposals have been considered for spacelike distance,
but all suffer a similar fate.  Spatial distance
in a discrete, Lorentz invariant context has unfamiliar subtle difficulties.
One can make
progress by breaking the Lorentz invariance, as we consider in section
\ref{Section 4}.  In section \ref{Section 5}, we replace the minimum above with an
average over suitably selected minimizing pairs, and present numerical
evidence that this is sufficient to overcome the difficulties described above.

\section{Details of simulations}\label{Section 3}

Here we present some simulations which show how 
the timelike and spacelike distance
measures described in section \ref{Section 2} perform `in practice'.
These simulations were all performed using a CausalSets toolkit
within the Cactus high performance computing framework
\cite{cactus}.  The Monte Carlo simulations are facilitated by a
MonteCarlo toolkit in Cactus, which provides a variety of methods
for generating independent parallel random numbers within the Cactus
framework.  For the computations described in this work, the random
numbers were generated via a multiple recursive linear congruential
generator proposed
by L'Ecuyer \cite{lecuyer}.\footnote{Note that there appears to be a
sign error in L'Ecuyer's implementation of his algorithm.  We use
the parameters specified in the main text of the article, rather
than those used in his implementation.}

In all our simulations we generate a causal set by sprinkling into some
finite convex region of Minkowski space.  Such a causal set obviously comes
with a faithful embedding into that same region.  We use this embedded causal
set to test the various measures of timelike and spatial distances described
in the paper.

In order to select elements of the causal set between which we
seek to measure distances, we simply select a number of \emph{target points}
in $\M^d$, and choose the causal set elements which are nearest the target
points, using a (flat) Euclidean metric.\footnote{The more typical procedure
  in this regard is to condition on the presence of causet elements at the
  target points (\eg at the endpoints of a causal interval).
  We instead choose to sprinkle without such conditions.
}
We should stress that this procedure is of no fundamental
importance, which is why we are justified in using a 
Euclidean metric, which is frame dependent, and also using the embedding in
such an explicit way.  It is simply a
procedure 
for selecting, in a practical and easy manner, elements from 
the causal set
on which we can employ our distance measures.
As the sprinkling density goes to infinity, the continuum distance between a
target point and the location to which the 
nearest element is mapped in the faithful embedding goes to zero.

For some of our simulations we do not use a proper Poisson
sprinkling, in that we hold the total number of sprinkled elements $N$ fixed.
For large $N$ this very closely approximates a true Poisson sprinkling.  In
fact, for all the values of $N$ used in this paper, the results for a fixed $N$
sprinkling are indistinguishable from those for a full Poisson sprinkling.
(This is partially due to the fact that our fluctuations are greater for
smaller causets, and these greater fluctuations overwhelm any effect due to
fluctuations in the total causet size.)
In other simulations we allow the total number of elements to vary
with respect to a Poisson distribution, which is characterized by some mean
value $\langle N \rangle$.

The results displayed in figures \ref{timelike2d}, \ref{m2eff}, \ref{bgdist2d},
\ref{bgdist3d}, \ref{obtuse_triangle-voldist}, and \ref{acute_triangle-voldist}
all come from sprinklings with a fixed total number of elements.  Those
displayed in figures \ref{timelike3d}, \ref{bgfailure}, \ref{2linkdist},
\ref{bgfailure-2linkdist}, and \ref{neighbors} come from genuine Poisson
sprinklings.  In figure \ref{m3eff} we use a fixed $N$ sprinkling for all but
the largest three data points shown; for those three the indicated value of $N$
represents the mean of the Poisson distribution.
Our results are split in this way simply because we updated the `sprinkling
engine' of the CausalSets toolkit during the course of our simulations for
this paper, primarily to check if it had any affect on our results.
Where the size of the causal set has been randomized, it is to be understood
throughout that $N$ (and $N_\diamond$ etc.) stands for the mean of the
appropriate Poisson distribution, rather than a
value from one sampling of it.

Most all results we present are from a Monte Carlo simulation in which we
generate causal sets `one off', and compute the distance measures as
described.  In the case that the distance measure is undefined for the
particular randomly generated causal set, we simply discard it from our
statistics (though we still count it against the total number of causets to
generate, so this occurrence will tend to lead to larger error bars).  This
can occur because the selected antichain does not have any elements it its
common future or past, or does not possess any future 2-links (see below).

\subsection{Timelike Distance}
\label{timelike_numerics}

For starters we consider (finite density) sprinklings into an
interval of $\M^2$.  Note that \whp{} elements of the causal set will
not get mapped to the endpoints of the interval.  We instead use the
endpoints of the interval as target points, so we use the closest elements in
the sprinkling to define an interval $[x,y]$ within the causal set.
Here we could have taken a different approach, as in \rf\cite{bg},
in which one 
conditions on there being causal set elements at the
`endcaps' of the interval. 
We prefer this approach because it more closely resembles the computations we
do later for various measures of spatial distances, without imposing any
condition on the embedding.

Here let us highlight why we should not use Lorentzian distance
when finding the elements closest to the target points (end-points
in this case). Imagine in 1+1-dimensions an interval, starting from
$(t=-1, x=0)$ (call it point $A$) and finishing at $(t=+1,x=0)$ (point
$B$). If we use Lorentzian distance to find the element closest to
$A$, it may turn out to be an element close to $(0,-1)$, since this
point is close to the lightcone of $A$. Similarly we may get for
point closest to $B$ an element close to $(0,+1)$. However, these
`nearby' points not only do not give us a good idea of the longest chain
in the interval between $A$ and $B$, they are not even
related in the first place! 
This is 
avoided 
by using the Euclidean metric.

In the following we use $N_\diamond$ to refer to the mean cardinality of $[x,y]$, and
$V_\diamond$ to refer to the spacetime volume of the interval in the continuum
$[\phi(x),\phi(y)]$.
$N$ and $V$ correspond to the (mean) total number of elements in the interval we
sprinkle, and the total spacetime volume into which we sprinkle,
respectively. As $N\rightarrow\infty$ the ratio
$\frac{V\diamond}{V}=\frac{N_\diamond}N\rightarrow 1$.

We wish to use length (number of links) of the longest chain $L$ 
between the 
elements $x$, $y$ to estimate the timelike geodesic separation $T_c$ between
the corresponding points in the manifold $\phi(x), \phi(y)$.  Using the theorem from
reference \cite{bg}
\be
\frac{L}{N_\diamond^{1/d}} \to m_d \;,
\ee
in probability in the
asymptotic limit $\rho\rightarrow\infty$, and the fact
that the volume $V_\diamond$ of a spacetime interval in $d$
dimensions is given by
\be
V_\diamond = D_d T^d
\ee
(where $D_d$ is
an $O(1)$ constant factor, in particular for our purposes we will need $D_2
= \frac{1}{2}$ and $D_3 = \frac{\pi}{12}$), one can estimate the
timelike separation in the manifold by
\beq
T = \frac{L}{m_d \rho^{1/d} D_d^{1/d}} \;.
\label{asymptotic_timelike_dist}
\eeq
This is of course valid only in the 
limit $\rho\rightarrow\infty$.  For finite $\rho$,
(\ref{asymptotic_timelike_dist}) consistently underestimates the continuum
distance $T_c$.  We refer to such underestimation as \emph{\tiunder}.  (In
fact, to leading order, in 1+1 dimensions, this underestimation is known to
fall off as 
one over the variance in $L$).

For the relatively miniscule sizes of causal sets which can be put on a
computer (\eg $N \ll 10^{64}$),
the deviation for finite $\rho$ is important.  To account for
this, we replace the ratio $m_d$ by an `effective ratio' 
\beq
\label{meffective} \meff{d} = \frac{L}{(\rho V_\diamond)^{1/d}} \;.
\eeq
As we can easily see, in the limit $\rho\rightarrow\infty$,
we have $\meff{d} \rightarrow m_d$.
Ideally we would like a functional
form of an
effective $m_d$ which would behave correctly at all scales. This
has not been calculated analytically (though much is known for the 2-dimensional case). Here we compute $L$ numerically and estimate
the functional form of $\meff{d}$ by curve fitting. 
In particular, for each sprinkled causal set, we compute the length of the longest
chain between our `nearest pair' $x$ and $y$, and from that compute the
quantity $\meff{d}$ from (\ref{meffective}).

\begin{figure}[hbtp]
  \psfrag{deviation}{$\left<\frac{T - T_c}{T_c}\right>$}
  \psfrag{log2 T}{$\log_2 T_i $}
  \includegraphics{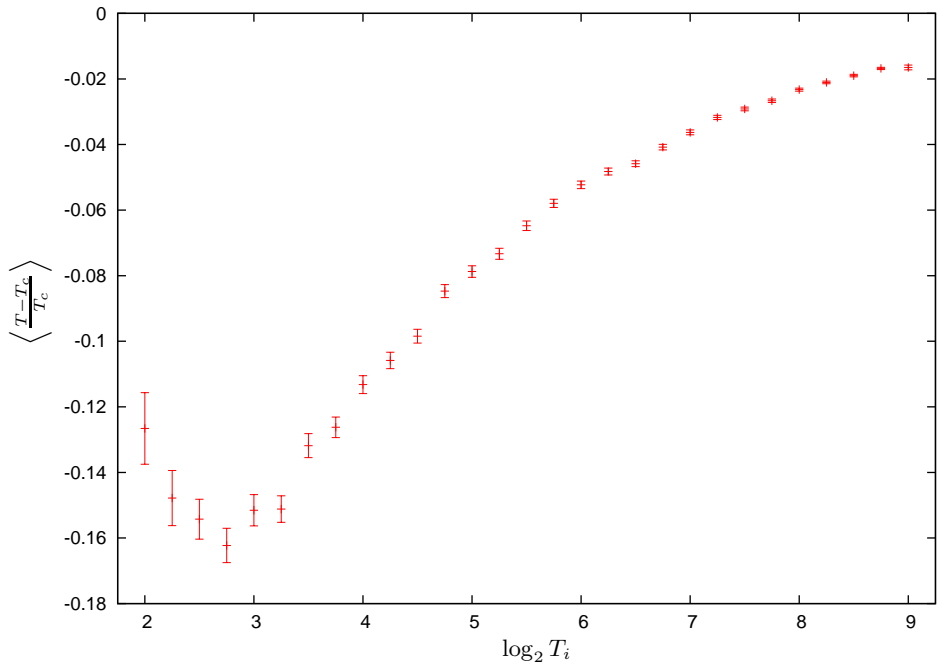}
  \caption{Measurement of proper time between a pair of elements in a causal
    set sprinkled into variously sized intervals of $\M^2$, by counting the
    length of the longest chain.  Each data point plotted indicates the mean
    deviation along with its standard error, for 900 runs.} 
  \label{timelike2d}
\end{figure}
In figure \ref{timelike2d}, we test how well the length of longest chain
approximates the timelike distance, in $\M^2$, if we use the known asymptotic
value of $m_2=2$. 
In particular, we perform a sequence of simulations, each of which sprinkles
900 causal sets into an interval of fixed height $T_i$, in fundamental units.
For each value of $T_i$, we plot a mean deviation $\left<\frac{T - T_c}{T_c}\right>$,
where $T$ is as given in (\ref{asymptotic_timelike_dist}), and $T_c$ is the
continuum distance $\sqrt{-(\phi_t(y)-\phi_t(x))^2 + (\phi_x(y)-\phi_x(x))^2}$
(where the subscript $t,x$ on $\phi()$ indicates the time or space coordinate
respectively).
In addition we include the error bar for each mean, which we estimate in the
usual manner from the sampled data values.
As one expects, the mean deviation goes to zero as $T\to\infty$. We also note
that for small distances, we observe considerable underestimation (the
phenomenon we call \tiunder). 
In fact, even for relatively `large'
distances that are easily representable on a typical workstation, we still get
a mean deviation of about 1.8\% from the continuum proper time.

In order to show that our distance measures overcome the difficulty presented
in section \ref{bgfailure_description} (and to avoid the comparatively large
computational effort of simulations in 3+1 dimensions), we will perform most
of our simulations in 2+1 dimensions.  Since analytic results on the value of
$m_3$ are restricted to bounds, and there are no results on the functional
form for $\meff{3}$ \cite{bachmat}, we will need to compute these quantities
numerically.  For `practice', we do this as well in 1+1 dimensions, where we
know both the exact value of $m_2$ and the asymptotic form of $\meff{2}$.

\begin{figure}[hbtp]
  \psfrag{log2 N}{$\log_2 N$}
  \psfrag{m2eff}{\hspace{-7mm}$\left<\meff{2}\right>$}
  \psfrag{g(x)}{\hspace{-7mm}$f(N)$}
  \psfrag{or(x)}{\hspace{-7mm}$f_{OR}(N)$}
  \includegraphics{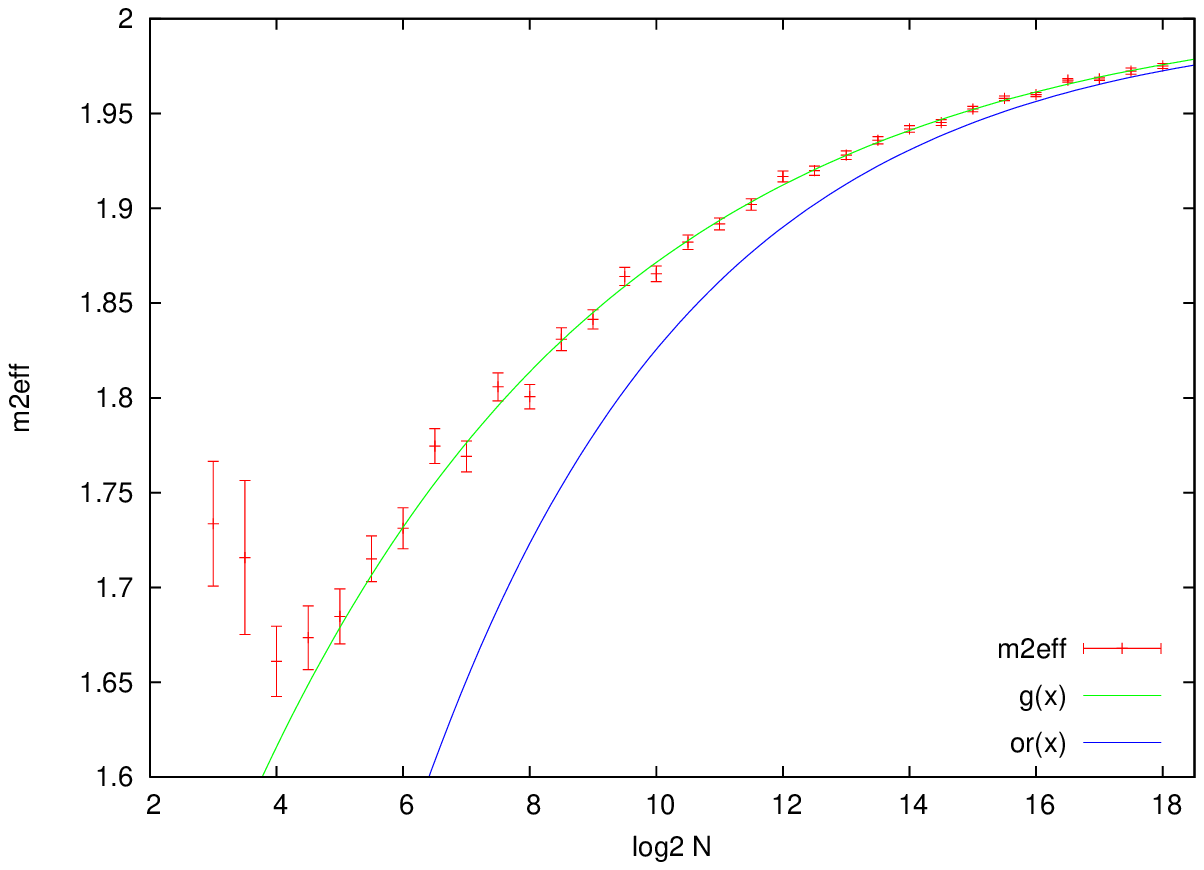}
  \caption{Correction to (\ref{asymptotic_timelike_dist}) for finite causets.
    Here we measure $\meff{2}$ by comparing with the known continuum timelike
    distance.
    $f_{OR}(N)$ is the leading order correction to the asymptotic value
    $\meff{2} = 2$ \cite{bdj}.}
  \label{m2eff}
\end{figure}
In Figure \ref{m2eff} we plot various values of $\left<\meff{d}\right>$ for
dimension $d=2$, for 400 runs of the same simulations as described above,
along with their standard error bars.
Here we use the total number of elements in the causal set $N = T_i^2/2$ for the
horizontal axis labels.  The asymptote is, as stated in ref.\
\cite{bg}, equal to 2.  We fit the data points for $N \geq 2^6$, to the function
\beq
f(N) = m_d + aN^{b/\ln 2} \;.
\label{fit_function}
\eeq
The parameter values are $m_2 = 2.0102 \pm 0.0037$, $a = -0.791 \pm 0.041$, and $b
= -0.1741 \pm 0.0077$.  The errors are as output from gnuplot's fit
command, which also gives $\chi^2 = 28.43$. Since it is not of central significance to the rest of the paper, we will not enter into the details of the curve
fit, apart from getting a numerical estimate of the value of $m_3$.
Here we see that we are able to reproduce the correct value of $m_2$ to within
a half of a percent.

The functional form we guessed to fit our data (\ref{fit_function}) is exactly
that of Odlyzko and Rains, as given in reference \cite{bdj}, with
$a \approx -1.758$ and $b = -\frac{\ln 2}{3} \approx -.231049$ (and $m_2=2$).
We include this function in figure \ref{m2eff}, for reference.
It appears that we are
able to reach the asymptotic approximation, at the largest causal sets we
have simulated.  Note that any difference between our procedure and those of
\cite{bdj, bg} is washed out for the largest causets.

We are now in a position to repeat the analysis in 2+1 dimensions.
\begin{figure}[hbtp]
  \psfrag{m3eff}{\hspace{-3mm}$\left<\meff{3}\right>$}
  \psfrag{g(x)}{\hspace{-4mm}$f(N)$}
  \psfrag{log2 N}{$\log_2 N$}
  \includegraphics{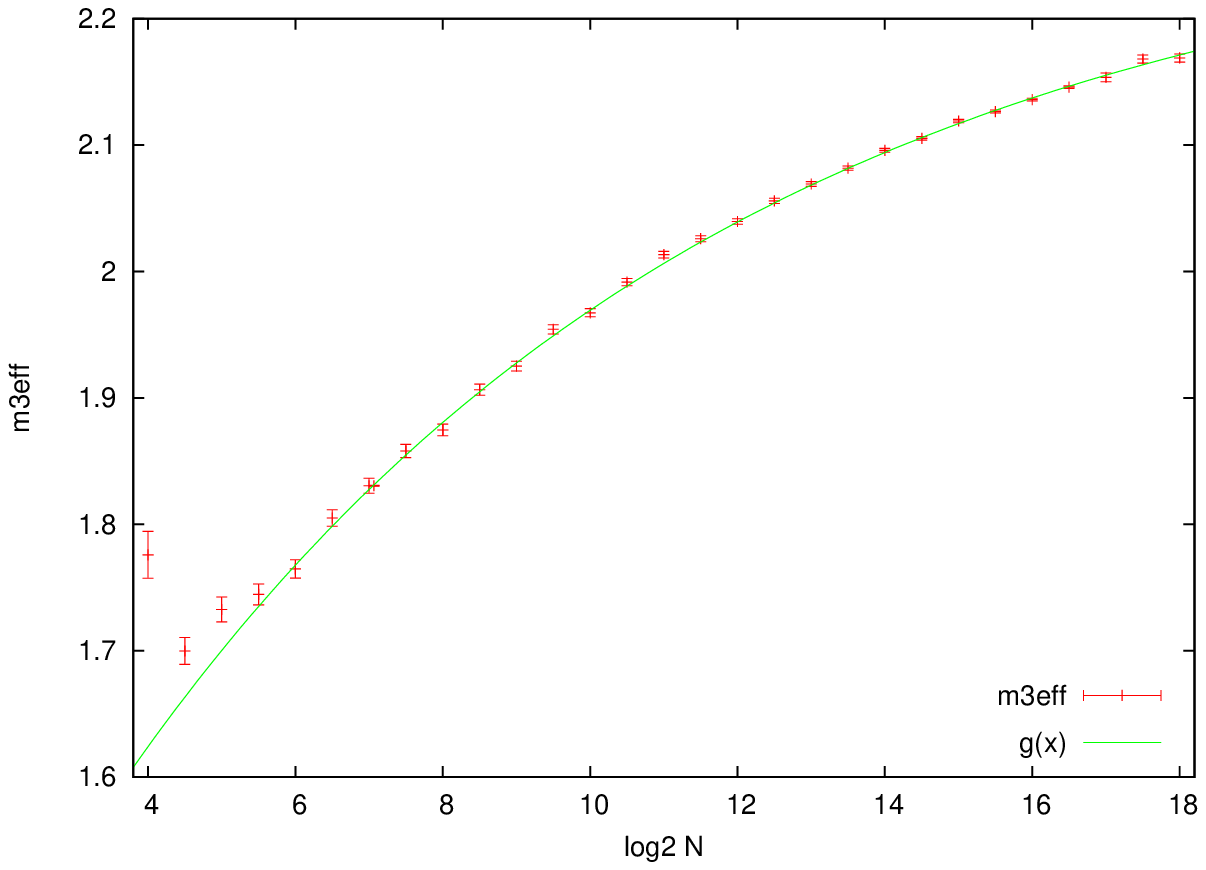}
  \caption{Convergence of length of the longest chain to proper time, in an
    interval of $\M^3$.  The fitting function shown is $f(N) = m_3 + a*e^{b
      \log_2 N}$.}
  \label{m3eff}
\end{figure}
In Figure \ref{m3eff} we again plot $\left<\meff{3}\right>$ for sprinklings
into an interval of $\M^3$.  For the smaller simulations ($N<2^{17}$), we
generate 1600 causets for each value of $N$, while for the larger simulations
we use fewer (400 at $2^{17}$ and $2^{17.5}$, 100 at
$2^{18}$).
We again fit to the functional form (\ref{fit_function}), and find the
parameter values $m_3 = 2.296 \pm 0.012$, $a = -1.087 \pm 0.014$, and
$b = -0.1201 \pm 0.0053$.
For almost all computations of a timelike distance using the length of the
longest chain in this paper, we will use such a numerical approximation to
$m_3$.\footnote{\label{wrong_m3}In practice we will sometimes use a value we computed earlier,
$m_3\approx2.278$, rather than the value above.  Since the corresponding
computations are only used to obtain qualitative results, the resulting 2\%
difference in timelike distances will have no affect on the conclusions of the
paper.}
An exception is in demonstrating the failure of \bgdist{} in section
\ref{bgfailure_sec}, where we will need an acurate numerical measurement of
timelike distance.

\begin{figure}[hbtp]
  \psfrag{deviation}{$\left<\frac{T - T_c}{T_c}\right>$}
  \psfrag{log2 T}{$\log_2 T_i$}
  \includegraphics{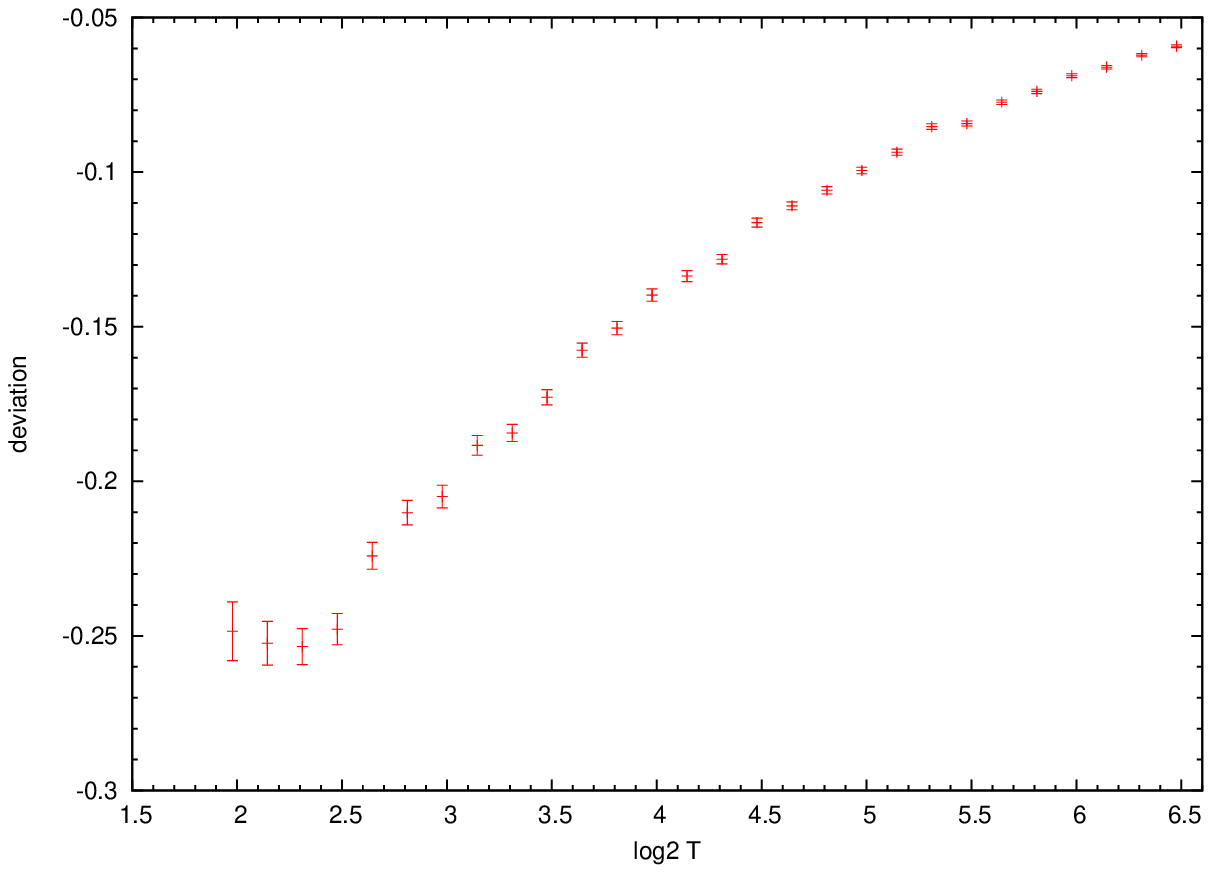}
  \caption{Measurement of proper time of a sprinkled interval in $\M^3$, of
    various sizes.}
  \label{timelike3d}
\end{figure}
In figure
\ref{timelike3d} we test how measuring length of the longest chain
differs from the continuum proper time in 2+1 dimensions, as we saw above in 1+1. In order to generate this plot we have
used the asymptotic value of $\meff{3}$, \ie $m_3 \approx 2.296$.
Note that there is still a $6\%$ deviation from the continuum distance, for
the largest causets we simulate.

\subsection{``\Bgdist{}'' and its failure} 

In this section, we will examine the \bgdist{}, and demonstrate that the
problem described in section \ref{bgfailure_description} is observable in
computer generated causal sets.
We recall that \bgdist{} was a successful
distance function for $\M^2$. However, in higher dimensions, it
failed, 
due to the existence of an infinite number of independent minimizing
pairs.  In computer simulations, it is only possible to sprinkle into regions
of finite volume, and therefore one can only consider causets which faithfully
embed into
a finite region of 
Minkowksi space. 
In a such finite region, 
there are only a finite number of \minpair{}s, and thus one cannot directly
observe
the failure of \bgdist{} (giving a trivial result).  However, one \emph{can}
observe some underestimation of the distance.
Here we demonstrate the latter, by considering a sequence of larger and larger subregions
of $\M^3$, and observe the \bgdist{} decaying toward its trivial result.

In order to avoid confusion, we should stress here that we will have two
different kind of plots. In section \ref{bgdist_sec}, we will consider how
\bgdist{} is performing when we increase the density of sprinkling, while keeping the
target points fixed. This means that the plots correspond to distances that
vary in fundamental units (as in previous sections), and some effects will
vanish in the limit of infinite distance in fundamental units.  In section
\ref{bgfailure_sec}, however, we will keep the distance in fundamental
units fixed, and only increase the size of the sprinkling region.  This means
that all the points of the plot in section \ref{bgfailure_sec} correspond to a
single point in the other graphs, in particular 
to the
point in which the distance between target points is 8 in fundamental units.
This is why we should \emph{not}
expect to see the effect of \tiunder{} to vanish (and thus we need an accurate
value for $\meff{3}$). Similarly, some other effects that vanish in the limit
in which the distance goes to infinity in fundamental units must be taken
into account in section \ref{bgfailure_sec}.

\subsubsection{Numerical results for \bgdist}
\label{bgdist_sec}

We first consider how \bgdist{} performs in $\M^2$ and $\M^3$,
for a fixed region in Minkowski spacetime, when sending the sprinkling density $\rho$
to infinity.\footnote{Equivalently we consider increasing
distances in fundamental units.}  Because the spacetime volume of the region is fixed, although
the number of minimizing pairs will diverge with $\rho$, we expect the number
of \emph{independent} such pairs to remain roughly fixed.  This is because, in
order for two minimizing pairs to be independent, their corresponding
intervals must have small overlap.  Given that the spatial separation between
the elements $x$ and $y$ are fixed, and the volume of the sprinkling region is
fixed, there is only room for so many `independent intervals', and thus the
effective number of minimizing pairs remains bounded.
Thus we expect that \bgdist{} will perform reasonably well in this scenario of
finite volume but diverging density.

\begin{figure}[hbtp]
  \psfrag{deviation}{$\left<\frac{X - X_c}{X_c}\right>$}
  \psfrag{log2 N}{$\log_2 N$}
  \includegraphics{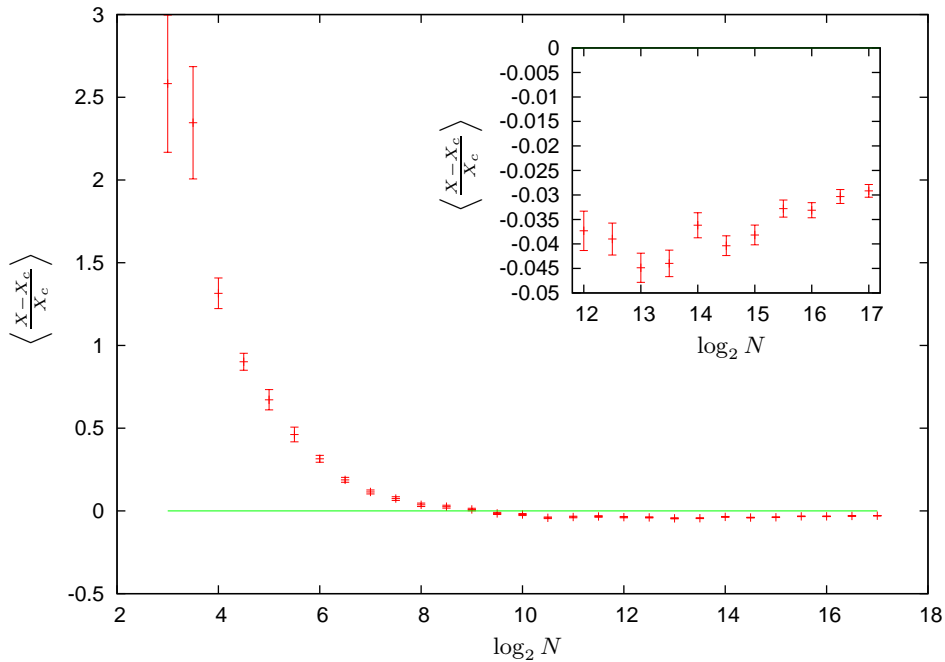}
  \caption{Measurement of \bgdist{} in 1+1 dimensions.  We sprinkle $N$
    elements into the
    Cartesian coordinate square $[-1,1]^2$.  Target points are at $(t=0, x=\pm
    \frac{3}{10})$.
    The inset plot zooms in on the large $N$ data, to show that the deviation
    is not constant, but behaves in a manner consistent with a drift toward zero.}
  \label{bgdist2d}
\end{figure}

In figure \ref{bgdist2d} we show the results of a sequence of simulations of
400 sprinklings into a fixed square of $\M^2$, at various densities.  A point
in the square takes coordinate values $t$ and $x$ in $(-1,1)$.  We choose
target points at $(t=0, x=\pm \frac{3}{10})$, and compare the spacelike
distance $X$ between
the corresponding causet elements as
described in section \ref{bgfailure_description}, with the continuum distance
between their embedding locations $X_c$.

The first thing to note is that for small $N$ 
we encounter a significant overestimation, that quickly vanishes at larger
$N$.\footnote{Small $N$ means equivalently
small distance in fundamental units.
}
A possible reason for this
is the fact that there will not exist causet elements exactly on the intersection of
the relevant lightcones.  The spacelike distance prescription selects
minimizing pairs which are more distant than these lightcone intersections,
thus leading to a larger estimate for spacelike distance.
We will refer to this phenomenon as 
\emph{\spover}.  Once we get to sufficiently large $N$ that \spover{} is
negligible, then we observe (solely) the now familiar \tiunder.
One may be concerned that it looks as if the deviation approaches a constant
negative value.  However, upon zooming into the large $N$ data points, we see
that the behaviour of the ratio is consistent with a drift toward zero, as one would expect.  The slowness
of the drift is due to the slow convergence of length of the longest chain to
the timelike distance.

\begin{figure}[hbtp]
  \psfrag{deviation}{$\left<\frac{X - X_c}{X_c}\right>$}
  \psfrag{log2 N}{$\log_2 N$}
  \includegraphics{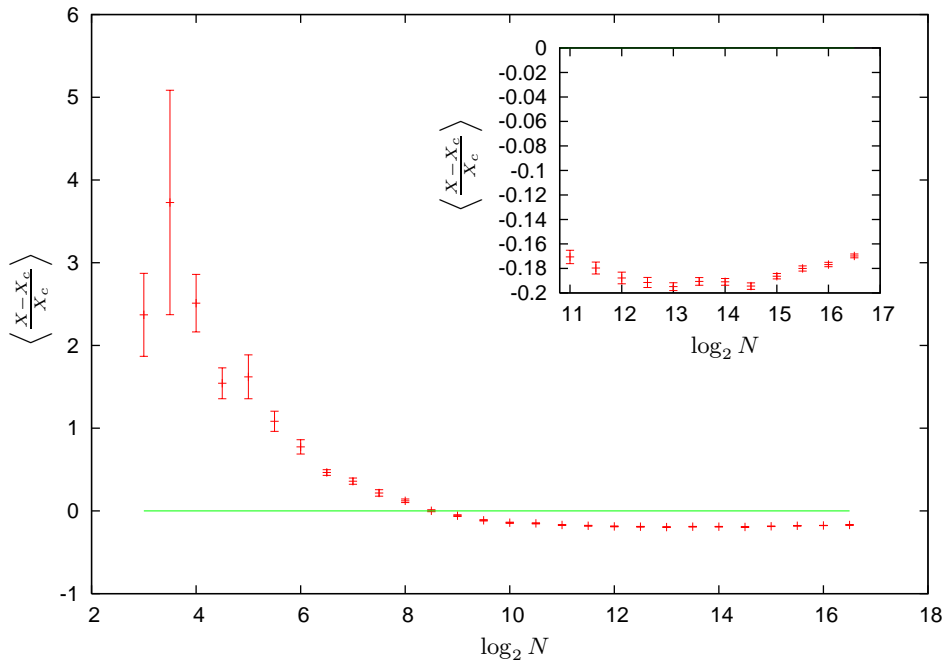}
  \caption{Measurement of \bgdist{} in 2+1 dimensional Minkowski spacetime.
    We sprinkle into the Cartesian coordinate cube $[-1,1]^3$, at varying
    densities.  Target points are at $(t=0, x=\pm \frac{3}{10}, y=0)$.
    The inset plot zooms in on the large $N$ data, to show that the deviation
    is not constant, but is consistent with a slow drift toward zero.}
  \label{bgdist3d}
\end{figure}

In figure \ref{bgdist3d} we show the analogous results for 400 sprinklings
into a fixed cube of $\M^3$.  The target points are $(t=0, x=\pm \frac{3}{10},
y=0)$.
Here we use the (older) asymptotic value
$m_3 \approx 2.278$.  The plots show the same features as in 2
dimensions, namely the existence of \spover, \tiunder, and also the fact that
the deviations seem consistent with a slow drift toward zero in the asymptotic limit.

\subsubsection{Demonstration of failure of \bgdist}
\label{bgfailure_sec}
Recall that \bgdist{} fails due to the existence of a large
number of minimizing pairs, whose corresponding intervals have small mutual
overlap (\ie{} independent minimizing pairs).
In order to be able to ``detect'' the failure of the \bgdist{} we
need to
perform simulations
which consider an increasing number of independent minimizing pairs.
To achieve this we will need to sprinkle into a very
large region.
Furthermore we have
noticed that there exists some underestimation from counting the length of the
longest chain for
timelike distance, which we called \tiunder. This underestimation is
corrected by using a measured value for $\meff{3}$, rather than the asymptotic value $m_3$.
There is also some overestimation that we have called \spover{} (the
effects of which explains the high values of the first points of
figures \ref{bgdist2d} and \ref{bgdist3d}).

With these issues in mind, we choose to sprinkle 900 causal sets into a box shaped region
of $\M^3$, with coordinate values $t \in (-T,T)$, $x \in (-4,4)$, and $y \in
(-T,T)$.  The two target points are at $(t=0, x=\pm 4, y=0)$.  We again
measure the spatial separation $X$, and compare with the continuum value
$X_c$.  As $T$ gets larger, we will capture increasingly longer portions of
the intersection of the light cones, near which the minimizing pairs lie.
Here we have chosen coordinates such that the sprinkling density $\rho$ is
fixed at 1. 
Thus the distance to be measured remains constant in fundamental units.

To overcome \tiunder{}, as mentioned above, we must carefully measure the
relevant value of $\meff{3}$.  Since the target points are separated by a
distance of 8, we expect the resulting minimizing pairs to be separated by a
distance of approximately 8.  The corresponding interval then has a volume of
$\frac{\pi}{12}8^3 \approx 134$.  A careful (averaging over 250 000 causets)
measurement of $\meff{3}(134)$ yields $1.83050 \pm 0.00046$.
\begin{figure}[hbtp]
  \psfrag{deviation}{$\left<\frac{X - X_c}{X_c}\right>$}
  \psfrag{log2 T}{$\log_2 T$}
  \psfrag{fixed N}{\hspace{-5mm}fixed $N$}
  \psfrag{0}{\hspace{-5mm}0}
  \includegraphics{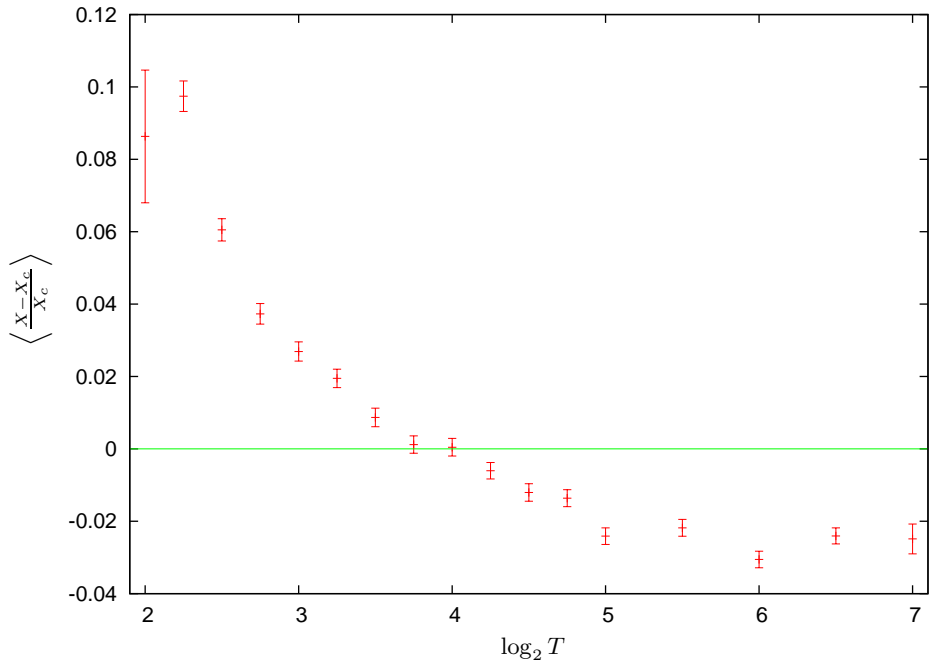}
  \caption{Measurement of spatial distance for a pair of elements at fixed
    separation, while considering sucessively larger regions of $\M^3$.
    The sprinkling region has coordinates $t \in (-T,T), x \in (-4,4), y \in
    (-T,T)$, with target points at $(t=0, x=\pm 4, y=0)$.  The fact that the
    distance is not constant (decreases) and moreover crosses below zero shows
    that we are seeing the failure of the \bgdist, due to the existence of
    many minimizing pairs and the random fluctuations of the sprinkling
    generated causal set.}
  \label{bgfailure}
\end{figure}

Since it is a small distance (8 in fundamental units), and the \tiunder{} has
been eliminated by using $\meff{3}$, we naively would
expect to get a positive value for the deviation, due to \spover.
However, in figure \ref{bgfailure}, we see two things that stress the
failure of the \bgdist{}. First, 
the value of the distance is \emph{not} constant, but by merely considering
successively larger regions of Minkowski space, we see its value reduce. The
second is that, even though we have no reason to get a value below
zero (\ie{} an underestimation), we see that even within our
computational powers, the value goes (systematically) negative.
Therefore, taking
into account more and more minimizing pairs not only balances the
(relatively large for small distances) \spover, but even produces some
underestimation.  In any event, this simulation 
confirms the well understood theoretical reason for the failure of
the \bgdist.\footnote{If we could send $T$ to infinity, we would expect to
  see $L \to 2$, so $X = \frac{\left(\frac{12}{\pi}\right)^{1/3} L}{\meff{3}}
  \to \sim 1.708$, and thus the deviation should converge to $(1.708 - 8)/8
  \approx -.787$.  To achieve the $L=2$ we would need a void in the
sprinkling of volume $\sim 8^3$.  The probability for this to occur for any
given region of this volume is $e^{-512}$.  Thus a rough estimate for the
lower bound on the size of a sprinkling needed to reach $L=2$ is $N=512 \, e^{512}$,
which is obviously far beyond our computational powers.}

One potential loophole which may occur to the reader is that we may have used too
small of a value of $N$ (134), when computing the appropriate $\meff{3}(N)$,
since the minimizing pairs will not fall exactly on the intersection of the
light cones of $x$ and $y$, but will be offset slightly from it (due to \spover).  Thus one may
expect that we should use $\meff{3}(N))$ for some larger value than 134.  A
glance at figure \ref{m3eff} reveals that this would give a larger value for
$\meff{3}$, and thus smaller values for $X$, which would have the affect of
shifting the datapoints of figure \ref{bgfailure} down by some amount.  This
merely strengthens the conclusion that we are seeing the failure of \bgdist{} in
our simulations.\footnote{One could also worry that, since the target points
  are at the edge of the sprinkling region, we will in fact see intervals with
  \emph{smaller} $N$.  We have checked for this explicitly 
  by adding causet elements at the
  target points by hand, and found that this does not change the fact that we
  consistently see deviations less than zero, for large $T$.}

\section{$n$-links, sphere distance and proximity}\label{Section 4}

So far we have explored earlier attempts at defining a spatial distance
measure on a causal set.
In this section we proceed with some new considerations.
Let us start with two definitions, the first in the continuum, and the second
on a causal set.\\
\textbf{Definition 1:} The common future/past of a collection of $n$ mutually
spacelike separated points $\{x_i\}$ is
\beq
J_c^\pm(\{x_i\})=\bigcap_{i=1}^n J^\pm(x_i)
\eeq
\\\textbf{Definition 2a:} In a causal set, a \emph{past $n$-link} of an $n$-element
antichain $x_1,x_2,\cdots,x_n$ ($n$ mutually unrelated elements)
is an element $y$ such that $y\prec x_m$ and $\nexists z$ such that
$y\prec z\prec x_m$, for all $1\leq m\leq n$.
\\\textbf{Definition 2b:} Similarly we define a \emph{future $n$-link} of an
$n$-element antichain $x_1,x_2,\cdots,x_n$ to be an element $y$ such that
$y\succ x_m$ and $\nexists z$ such that $y\succ z\succ x_m$ for all $1\leq
m\leq n$.

\subsection{$n$-links and the dimension}

To gain an intuitive understanding of the meaning of these definitions, for
causal sets which faithfully embed into $\M^d$, first consider the definition of a link
in a causal set.  A link is an `irreducible connection' between a pair of
(related) elements of a causet.  By irreducible we simply mean that there are
no intervening elements, and thus a link represents the shortest possible
timelike distance between a pair of elements.  Naturally this corresponds to
the null relation in the continuum.  Thus the collection of causet elements
which are linked to a given element $x$ will tend to lie very close to (while
maintaining the causal relation) the past and future light cone of $x$.

Given this fact, one can intuitively understand the $n$-links as corresponding
to sprinkled elements which are within, and close to, the light-cones of all
$n$ unrelated (\ie{} spacelike) elements considered.  Let us first
explore 
the expected number of $n$-links of a given $n$-element antichain, in a causal
set that has arisen as a sprinkling into $\M^d$. 
We will consider separately the cases where (a) $n>d$, (b) $n<d$, and finally
(c) $n=d$. Note that we consider future $n$-links, but an identical analysis
applies for past $n$-links.

The
following equation yields the expected number of $n$-links in $d$
dimensions.
\be
\nlinks{n} = \rho\int_{J_c^+(\{x_i\})}dx^{d}\exp \{-\rho \:
\mathrm{Vol}(J^-(x)\cap[\cup_{i=1}^n J^+(x_i)])\} \;,
\ee
where $\mathrm{Vol}(R)$ indicates the spacetime volume of region $R$.
The expression is simply a sum of independent random variables,
each of which is the probability $dx^{d}$ of finding a sprinkled element at
$x$, times the probability that that element is linked to each of the $x_i$
(so the region causally between $x$ and the $x_i$ must be empty of sprinkled elements).
The region of
integration is the common future $J_c^+(\{x_i\})$.

As an illustrative example, we will consider the expected number of future
$2$-links in 1+1 dimensions ($\M^2$). (Here and throughout the paper, when we
speak of an expected number of $n$-links, it will always refer to the expected
number of $n$-links of a given $n$-antichain.)
Here we use fundamental units in which $\rho \equiv 1$.
We choose our 2-antichain with $y_1=(-r, -r)$
and $y_2=(-r, r)$\footnote{\label{caveat}
  Admittedly we are conditioning on there being sprinkled elements at these
  points in spacetime, which is a zero probability event.  However, we do not
  expect the results to change significantly if we instead use these points as
  target points, to select a pair of elements of the causet.}. We get the
integral
\be
\nlinks{2} = \int_0^\infty dt\int_{-t}^{t}dx\exp(-(t^2 + 4rt - x^2)/2)
\ee
since we subtract the overlap of the two regions
$[y_1,x]_c$ and $[y_2,x]_c$.\footnote{We will refer to causal intervals in the
  continuum by $[y,x]_c$, and in these expressions $x$ refers to a point in
  the domain of integration.}
By observing that, within the region of
integration, $t^2-x^2$ is positive and corresponds to the overlap of
$[y_1,x]_c$ and $[y_2,x]_c$ we conclude that
\be
\nlinks{2} \leq \int_0^\infty dt\int_{-t}^tdx\exp(-2rt) = \frac{1}{2r^2}
\ee
This was possible since, in order to be a $2$-link, the whole region
$[y_1,x]_c\cup [y_2,x]_c$ had to be empty. If we require a subset of this region
to be empty, we are guaranteed to get at least all the $2$-links, along with
some extra non-2-link elements. Therefore, this provides an upper bound on the
number of $2$-links. Note also that the above upper bound implies that the
number of $2$-links in $\M^2$, at most, falls off as $2/(2 r)^2$, as the
distance between the points ($2 r$) increases.

In $1+1$ dimensions we can easily see what happens to the expected
number of $1$-links (usually referred as links) and the $3$-links.
The former are infinite, as is well known.
The expected number of $3$-links will
decrease as the distance between the points increases (as do the
$2$-links) but much more quickly. 
For example, choosing points
$y_1=(-r,-r)$, $y_2=(-r,0)$ and $y_3=(-r,r)$, we get
\be
\nlinks{3} = \nlinks{2}\,\exp\{-(r)^2/2\} \leq \frac{\exp\{-( r)^2/2\}}{(4 r)^2}
\ee
which means that
the expected number of $3$-links drops exponentially faster than the
$2$-links, as the distance $r$ between the points increases.  By similar
arguments for higher dimensions, we arrive at the following table, where the
``$\rightarrow$'',
is understood as $r\rightarrow\infty$ (remember that $r$ is in fundamental
units).
\begin{center}
\begin{tabular}{|c|c|c|c|}
\hline
$\downarrow$n-link $\mid$ dim $\rightarrow$&1+1&2+1&3+1\\
\hline $\nlinks{1}$ & $\infty$&$\infty$ &$\infty$\\
\hline $\nlinks{2}$ & $<2/(2 r)^2\rightarrow 0$& $\infty$ & $\infty$\\
\hline $\nlinks{3}$ &$<\frac{\exp\{-(r)^2/2\}}{(4 r)^2}\rightarrow 0$ &$\rightarrow 0$ &$\infty$\\
\hline
\end{tabular}
\end{center}
These results generalize:
\begin{itemize}
\item[(a)] $n>d$. In this case, typically we have no $n$-links. This means
  that for $n$ unrelated elements, 
  the probability of existence of an $n$-link is very small, and decreases
  rapidly for elements which are far from each other in the embedding. The
  $n$-link corresponds to an element that is just inside the lightcone of all
  $n$ unrelated elements. 
  If $n>d$, and none of the elements are very close to each other, there is no
  point (in the continuum) that can be only an $\epsilon$ distance from the
  light cones of all $n$ elements. In the causal set, this is translated as a
  very small probability for existence of $n$-links.

\item[(b)] $n<d$. Here typically we will have infinite $n$-links (as for
  $1$-links in 1+1, or $2$-links in 2+1 dimensions). The reason is that in the
  continuum there exist a continuous infinity of points in the intersection of
  the lightcones, in particular a $(d-n)$-dimensional surface. Corresponding
  to each of these points there is a small but non-zero probability of a
  causal set element existing within a given, but arbitrarily small, distance.
  The infinity of such points in the continuum implies that there will be an
  infinite number of causal set elements close enough to give an $n$-link.

\item[(c)] $n=d$.
  In the continuum there exists a unique point that
  corresponds to the intersection of the $n$ lightcones. In the faithfully
  embedded causal set,
we expect there to be a non-zero but rather small probability (depending on
the separation distance of the elements), as we have seen in the table, for a
given $d$-antichain to possess a $d$-link.
\end{itemize}

Given these observations of the dependence of the behavior of $n$-links on
dimension, it seems reasonable that counting of $n$-links could be used as a
dimension estimator for causal sets.  The idea would be roughly to select any
$n$-antichain, for $n=1$ (a single element), and count the number of links
emanating from that element.  Then increment $n$ by one (select another
unrelated element) and repeat.  The first value of $n$ for which one does not
find a large number of $n$-links would be the dimension.  In practice one may
need to sample a number of antichains for each value of $n$ in some manner.

In fact one can imagine that this itself may be a rather stringent necessary
condition for a causal set to be `manifoldlike', \ie{} that it is likely to
faithfully embed into some smooth spacetime manifold.
If the numbers of $n$-links do not behave in the manner described above, then
the causal set is not so likely to have the light cone structure that one
expects from a continuum spacetime.  Such a principle could be insightful in
formulating a dynamics for causal sets, which yields faithfully embeddable
causets, `with high probability'.

\subsection{Equidistant points, `Sphere-distance' and $l_g$}
\label{diameters_sec}

As we have noticed earlier, the failure of the \bgdist{} was due to the
existence of an infinite number of ``minimizing pairs'', in higher
dimensions. Intuitively, this arises because in those higher dimensions, the
locus of points (in the continuum) that lie exactly \emph{on} both light-cones
(of spacelike points $\{x_i\}$) is \emph{not} a single point, but a full
hyperboloid (a co-dimension 2 submanifold), and moreover one of infinite
volume. The latter is important for the discrete case, since it would
correspond to an infinite number of elements, while if the surface had finite volume in the
continuum, it would correspond to only a finite number of elements of the
causet.

Consider that this infinity of minimizing pairs arises because of Lorentz
invariance.  If we select two spacelike points of $\M^d$, then we have
defined a system which remains invariant under
a $d-2$ dimensional subgroup of the Lorentz group, namely those boosts which
hold the pair of points fixed.
In order to construct a non-trivial distance measure on a causal set, this
suggests that we must select enough elements to eliminate this boost freedom
of the system, \ie{} we need a $d$-antichain.
Note that the future (and also past) light cones of $d$ spacelike points in
$\M^d$ intersect at a single point\footnote{In the frame of simultaneity this
  reduces to finding the intersection of a collection of $d-2$-spheres of
  equal radius.},
which suggests that the $d$-antichain may have a well
defined distance measure associated with it.  We consider two possibilities,
which correspond to measuring the diameters of the bounding and circumscribing
spheres defined by the $d$ points.

 First we notice that the distance between the unique point on the $d$ common
 future lightcones and the one on the past, corresponds to the diameter of the
 circumscribing\footnote{The $d-2$ sphere that has all the $d$ points on its
   boundary.} $d-2$ sphere.  Unfortunately, we do not have an intrinsic way to
 find the element closest to the $d$ common lightcones. (Here and throughout
 the paper, by the phrase ``$d$ common lightcones'', or simply ``common lightcones'', we mean the intersection of
 the lightcones of the $d$ spacelike points.)
If past and future
 $d$-links existed, we could use those, but as we have seen in the previous
 section, in $d$-dimensions, in general, we find \emph{no} $d$-links for an
 arbitrary selection of $d$ unrelated elements.

An obvious choice is to use a procedure like the one in the \bgdist{} to
locate this intersection point.  We make the following definition.\\
\textbf{Definition 3:} Given $d$ mutually unrelated elements $x_i$, we
define the \emph{\lgdist} $l_g(x_1,\cdots,x_d)$ in
$d$-dimensions, to be
\be
l_g(x_1,\cdots,x_d):=\min_{w,z} d(w,z)\quad \textrm{where}\quad w\in
  J_c^-(\{x_i\}) \quad \textrm{and} \quad z\in  J_c^+(\{x_i\})
\ee
However, minimizing over all
the common future/past does not always `select' the unique
point of intersection of the light-cones. If it did,
the distance recovered would be the diameter of the
circumscribing $d-2$ sphere. Instead, the \lgdist{} measures
the diameter of the smallest
$(d-2)$-sphere that \emph{contains} (in the interior or on the boundary) the $d$
spacelike elements (also known as ``bounding sphere''). These two
spheres coincide (in 2+1 dimensions) when the included angles at all the $d$
elements (vertices) are
acute,
and in this case the minimizing pairs will tend to lie close to
the intersection of the lightcones, so $l_g$ should work well.  In higher
dimensions, an analogous statement will be true, with higher dimensional
generalizations of the angles.  In the case that the spheres disagree,
however, a similar problem arises with the one in \bgdist{}, since we will
find many, though not infinite, minimizing pairs.  Since the minimizing pairs
are many but not infinite, we expect some underestimation of the distance,
though we should \emph{not} get the trivial result of \bgdist. 

To visualize the above, 
consider again figure \ref{minimizing_pair},
  though this time 
there is a third element, at the red dot,
  which for convenience lies in the plane of the page.  The projection of its
  lightcones with the page are drawn in red.  For the computation of \lgdist{} the
  minimizing pairs must also be related to this third element, so they are
  prevented from `climbing' infinitely far up or down the left side of the
  hyperbolae.  This makes the number of minimizing pairs finite.  It also
  seems reasonable that, unless this third element lies extremely close to the
  line segment connecting the first two, the number of minimizing pairs will
  not even be large, so \lgdist{} should still perform quite well.

To conclude, we can use the \lgdist{} to accurately estimate the diameter of
the bounding sphere, for unrelated elements that `form acute angles'\footnote{From
now on when we refer to `acute angles' it is understood, in the case of higher
dimensions, that the bounding sphere coincides with the circumscribing sphere.}.
In other cases, we will get some underestimation, though we expect it to be
extremely small in almost all circumstances. However, as of yet, we cannot
know a priori whether the elements form obtuse or acute angles.

~\\
If one is more careful, it turns out to be possible to select a pair of
elements which \emph{will} yield the diameter of the circumscribing sphere.
To describe the procedure, we first construct it in continuum Minkowski space.

We begin by defining our notation.  Each point 
in Minkowski space is represented by a vector of its Cartesian coordinates
\be
\x_i=(t^i,x_1^i,\cdots,x_{d-1}^i) \;.
\ee
The \emph{spacetime distance}
between any two points $\x_i,\x_j$ is defined to be
\be
(\x_i\x_j)^2 = (t^i-t^j)^2-(x_1^i-x_1^j)^2-(x_2^i-x_2^j)^2\cdots
  -(x_{d-1}^i-x_{d-1}^j)^2
\ee
If the distance is positive, the points are timelike separated, if
negative they are spacelike, while if it is zero they are null.

Let us consider $n$ spacelike separated points $\x_1,\x_2,\cdots,\x_n$.\\
\textbf{Definition 4 :} A point $\textbf{w}\in J_c^\pm()$ is 
future / past $n$-equidistant with respect
to the
$n$ spacelike points 
if
\be
(\textbf{w}\x_i)^2=(\textbf{w}\x_j)^2\quad\forall\quad i,j = 1 \ldots n \;.
\ee
We could have also define it alternatively (in fact equivalently in the
continuum) by 
requiring that the corresponding volumes of the Alexandrov intervals match.
The set of all such future /
past $n$-equidistant points is labeled $\Theta_n^\pm$. Note that if
$n>d$ we will typically find no $n$-equidistant points.

\noindent\textbf{Definition 5 :} We define the \emph{$n$-equidistance} of a
future / past $n$-equidistant point $\y$ with respect to $n$ spacelike points
$\x_1,\x_2,\cdots,\x_n$ to be the spacetime distance between the point $\y$
and any of the $\x_i$.
\be
l_{y}^2=(\y\x_i)^2 \;.
\ee
This of course is the same as the average over $i$ of the spacetime distances
$(\y\x_i)$.

Let us now derive an expression that relates past and future
$d$-equidistances with the diameter, $l_s$, of the $(d-2)$-sphere which
circumscribes $d$ spacelike points.\\ 
\textbf{Theorem:} Given $d$ spacelike points $\x_i$, in
$d$-dimensional Minkowski spacetime, one past $d$-equidistant point
$\p$ with equidistance $l_p$ and one future $d$-equidistant point
$\textbf{f}$ with equidistance $l_f$, 
the spacetime distance
$(\p\textbf{f})^2:=l_m^2$ goes as
\be 
l_m^2 = l_p^2+l_f^2+\frac{l_s^2}2+2\sqrt{(l_p^2+\frac{l_s^2}4)}
  \sqrt{(l_f^2+\frac{l_s^2}4)} \;.
\ee

This equation can be solved for the diameter $l_s$
\beq\label{l_s function of l_m}
l_s=(l_m^4+l_p^4+l_f^4-2l_p^2l_f^2-2l_m^2l_p^2-2l_m^2l_f^2)^{1/2}/l_m \;.
\eeq

\begin{proof}
We first choose coordinates such that all the $d$ spacelike points are
in the $t=0$ plane.  Moreover we choose the origin $\textbf{O}$ to
be at equal spacelike distance from each of the $d$ points, so that
$\x_i=(0,x_1^i,\cdots,x^i_{d-1})$, where $\sum_{n=1}^{d-1}
(x_n^i)^2 = l_s^2/4$ for each of the $\x_i$. 
From the
definition of equidistant point we have that the interval
$(\p\x_i)^2$ goes as
\be
(\p\x_i)^2=(t^p)^2-\sum_{n=1}^{d-1}(x_n^p-x_n^i)^2=l_p^2
\ee
for all $i = 1 \ldots d-1$.  From this we get that
\bea
\sum_{n=1}^{d-1}
x_n^p(x_n^i-x_n^j)=0& &\quad\forall\quad i,j \in [1,d-1]\nonumber\\
\p \cdot(\x_i-\x_j)=0& &\quad\forall\quad i,j
\eea
using that $\sum_{n=1}^{d-1} (x_n^i)^2=l_s^2/4$ for all $\x_i$'s.
The dot in the above expression is the usual Minkowski inner product.
Since the spatial components of the $d$
\spacelikepoints{} are arbitrary, 
almost certainly (\ie with
probability one) the $d-1$-surface at coordinate $t=0$ is
spanned by the $(d-1)(d-2)$ vectors of the form $\x_i-\x_j$. (For $d=2$ we
must replace the $(d-1)(d-2)$ vectors $(\x_i-\x_j)$ with the single vector $\x_1$.)
From this, it follows that
\be
x_n^p=0\quad\forall \quad n = 1, \ldots d-1 \;.
\ee
We now have that $\p=(t^p,0,\cdots,0)$, and by
similar arguments $\f=(t^f,0,\cdots,0)$. 
We also have the equidistance to be
\bea
l_p^2&=&(t^p)^2-l_s^2/4\Rightarrow\nonumber\\t^p&=&-\sqrt{l^2_p+l_s^2/4}
\eea
with the minus sign because it is \emph{past} equidistant, and similarly
for the future equidistant point (without the minus sign). We are
now in a position to express the interval $(\p\f)^2=l_m^2$ as a
function of $l_p,l_f$ and $l_s$.
\bea
l_m^2 & = & (t^p-t^f)^2\nonumber\\ & = &
  l_p^2+l_f^2+l_s^2/2+2\sqrt{l_p^2+l_s^2/4}\sqrt{l_f^2+l_s^2/4}
\eea
Since this final expression is in term of only invariant quantities,
it holds for all coordinates.
\end{proof}
It is now obvious that if we can define the above quantities for a
causal set, we can use (\ref{l_s function of l_m}) 
to compute the radius of the $d-2$-dimensional circumscribing sphere of
$d$ unrelated elements.

Before defining (past/future) $n$-equidistant elements, and $n$-equidistance,
for a causal set, let us note that $n$-links are essentially $n$-equidistant
elements, with zero equidistance.  They are by definition
(when they exist) the closest equidistant elements.\\
\textbf{Definition 6:} Given $n$ mutually unrelated elements of a causal set
$x_1,\cdots,x_n$, we define future $n$-equidistant
elements as the elements 
\beq
w \in \bigcap_{i=1}^n \fut(x_i) \quad\textrm{such that}\quad |[x_i,w]| =
|[x_j,w]|
  \quad\forall\; i,j = 1 \ldots n \;.
\label{equidist_elts}
\eeq
Past $n$-equidistant are defined equivalently, with $\fut(x_i)$ replaced by
$\past(x_i)$.  Note that here we have used \voldist, in place of the length of
the longest chain, to locate equidistant elements.  One can use the length of
the longest chain as well, but it is much less efficient in locating the
appropriate elements of the causal sets which give a good measurement of
$l_s$.\\
\textbf{Definition 7:} We define the $n$-equidistance $l^c_y$
of a (past/future) $n$-equidistant element $w$ with respect to $n$
unrelated elements $x_1,x_2,\cdots,x_n$ to be the \voldist{} between it and
any of the $x_i$ (all of which are equal, by definition 6).
\beq
(l_w^c) = |[x_i,w]|^{1/d}/D_d
\eeq
The superscript `c' will be dropped if there is no ambiguity.

One issue that is of concern is whether $d$-equidistant elements
exist in causal sets which are faithfully embedable in $\M^d$, with respect to
any $d$-antichain, and whether they are finite in number.
It can be shown (see Appendix \ref{proof of equidistant points}) that in 1+1
dimensions there are an infinite number of equidistant elements.
Based on this and on numerical
simulations for higher dimensions, we conjecture that $d$-equidistant elements
exist and are infinite in number, for any dimension.

Now we are now in a position to define the ``sphere distance'' associated with
a $(d-2)$-sphere that is circumscribed by $n$ spacelike elements.
The quantity is defined 
intrinsically to the causal set, and has the property that there is no significant
systematic over- or under-estimation at large distances.\\
\textbf{Definition 8:} Given $d$ mutually unrelated elements $\x_i$ of a
causal set, and any pair
of their past $\textbf{p}$ and future $\textbf{f}$ $d$-equidistant elements,
we define the \emph{sphere-distance}
$l_s(x_1,\cdots,x_d)$ 
to be given by (\ref{l_s
function of l_m}) where the quantities $l_m,l_p,l_f$ are computed using the
\voldist{} of (\ref{volume-interval}).\\
In numerical simulations we have used
the past/future equidistant elements with smallest equidistance in
order to compute $l_s$. However, in general one could use any pair of
equidistant elements
equally well.  In our simulations the value obtained matched very closely with
that deduced from the faithful embedding, \cf section \ref{diameter_numerics}.

\subsection{Numerical Results}
\label{diameter_numerics}

In this section we test the ``diameter measures''
$l_g$ and $l_s$ on
causal sets generated by sprinkling into an interval of $\M^3$. It is
fairly straightforward to generalize the calculation to higher dimension,
though the configurations considered would necessarily be more complicated,
and we expect none of the results to change.
In particular, we test $l_g$ and $l_s$ on
two `triangles' in $\M^3$, one obtuse (contains an included angle $>
\frac{\pi}{2}$) and the other acute (all included angles $<
\frac{\pi}{2}$). Here by a ``triangle'' we mean that we select three
points in $\M^3$ as target points for sprinkled elements, which lie
at the vertices of a triangle. Each of the triangles lie in the
$t=0$ plane.

Each simulation involves generating 100 causal sets, each with exactly the
number of elements $N$ as shown in figures \ref{obtuse_triangle-voldist} and
\ref{acute_triangle-voldist}.  Since it is possible that, for any given causal
set, one will not find both a past and future equidistant element, we
introduce the notion of \emph{equidistance quality}.  The equidistance quality
of a causet element $w$, with respect to a $d$-antichain $x_1, x_2, \ldots,
x_d$, is given by
\beq
\sum_{x_i} \left( |[w,x_i]| - \langle |[w,x_i]| \rangle \right)^2 \;,
\label{equidist_quality}
\eeq
where the sum indexes over the elements of the antichain $x_i$, and $\langle
|[w,x_i]| \rangle$ is the mean of the quantities $|[w,x_i]|$.
Note that for an (`exact') equidistant element $w$, this quantity will
vanish.  In our simulations we choose the closest\footnote{In practice we use
  the time coordinate of the embedding to decide which element is closest.
  Since, as stated earlier, it makes no difference which equidistant element
  one chooses, this should not affect our results.  In any event we have never
  observed the collection of equidistant elements, as a sub-causal set, to
  form anything other than a chain.  (Though we did not particularly look for
  this either.)}
element for which the
quantity (\ref{equidist_quality}) is as small as possible.
Note that this extra definition is needed for purely practical reasons, given
that we cannot sprinkle into the whole of Minkowski space.  It seems clear that
 one will always be able to find elements of
equidistance quality zero, \emph{somewhere} in a sprinkling of the entirety of
Minkowski space.

\subsubsection{Obtuse triangle}

\begin{figure}[hbtp] 
  \resizebox{10cm}{!}{\includegraphics{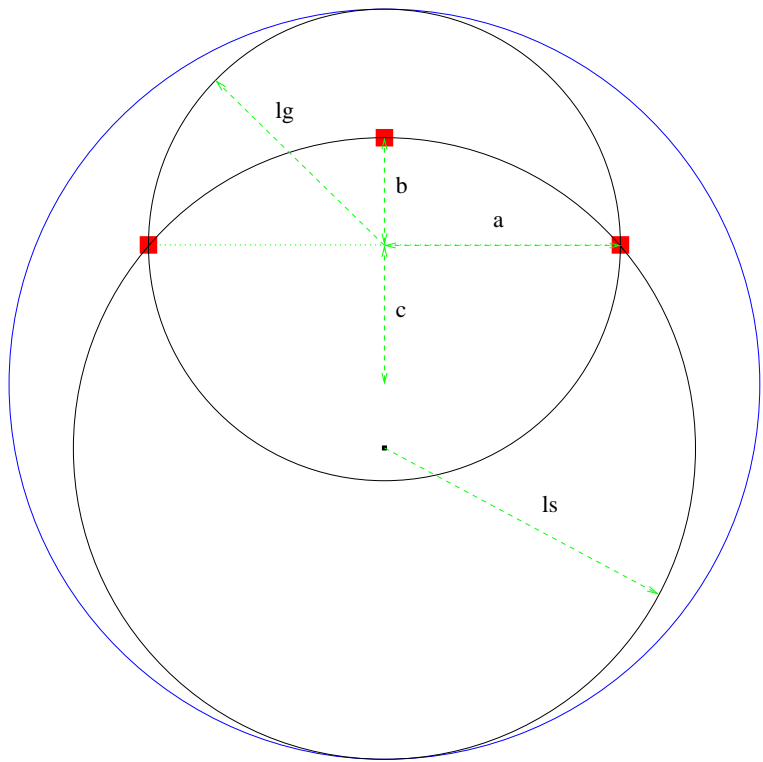}}
  \caption{Our obtuse triangle, in the $t=0$ plane.  In our coordinates $a=2.2$,
    $b=1$, $c=1.42$.  The causal set is sprinkled into the
    causal interval with endpoints $(t=\pm 6.5614, x=y=0)$.  The
    large blue circle is the intersection of the two lightcones which form
    the interval, and hence has radius 6.5614.
    ($c$ is the distance between the center this largest
    bounding circle, and the center of circle which bounds the three target
    points.)}
  \label{obtuse_triangle}
\end{figure}

\begin{figure}[hbtp]
  \psfrag{deviation}{$\left<\frac{l - l_c}{l_c}\right>$}
  \psfrag{lg deviation}{\hspace{-10mm}$\left<(l_g - 4.4)/4.4\right>$}
  \psfrag{ls deviation}{\hspace{-12.5mm}$\left<(l_s - 5.84)/5.84\right>$}
  \psfrag{log2 N}{$\log_2 N$}
  \psfrag{0}{\hspace{-5mm}0}
  \includegraphics{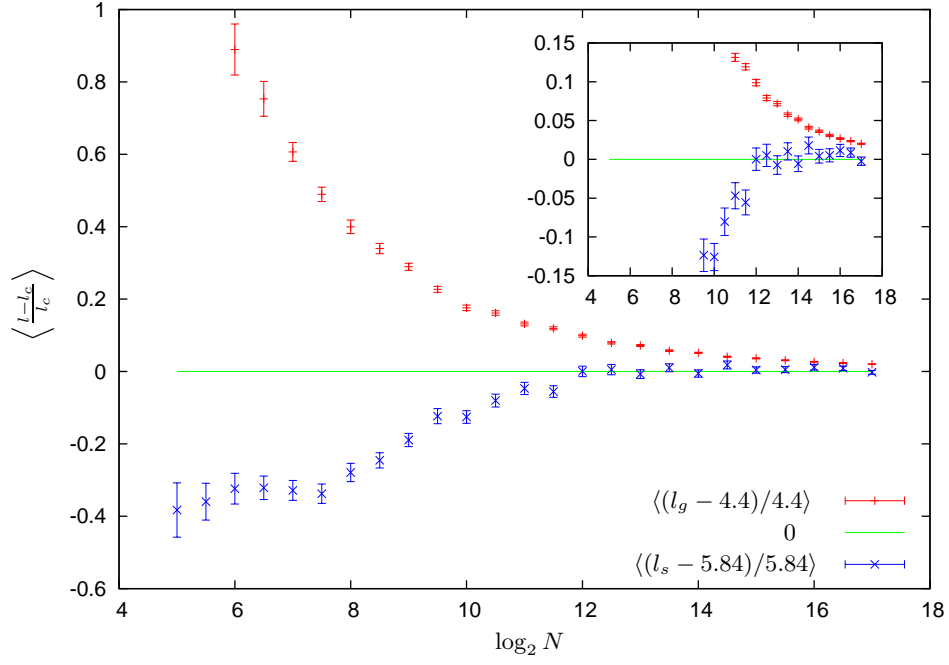}
  \caption{Performance of diameter measures for an `obtuse triangle', on
    causal sets sprinkled into an interval of $\M^3$.  On the vertical axis,
    $l$ stands for $l_g$ or $l_s$, while $l_c$ stands for 4.4 or 5.84,
    respectively.  The inset plot zooms in on the asymptotic behavior.}
  \label{obtuse_triangle-voldist}
\end{figure}

Figure \ref{obtuse_triangle} illustrates our obtuse triangle. For
this configuration the continuum values of the diameter of the
bounding and circumscribing spheres (circles in this case) are
respectively $l_g=4.4$ and $l_s=5.84$. Note that these values are
in arbitrary units, \emph{not} in fundamental units.  We consider successively
larger sprinkling densities $\rho$, while holding the target points fixed.
In figure
\ref{obtuse_triangle-voldist} 
we demonstrate the following results: $l_g$
converges (for large density $\rho$) to 4.4 as expected, while $l_s$
converges to the value 5.84.  The inset shows the asymptotic behavior in
greater detail.
For `small triangles' / small $\rho$ we observe the usual \spover{} in $l_g$,
and some small underestimation in $l_s$ (for whatever reason).

\subsubsection{Acute triangle}

\begin{figure}[hbtp]
  \resizebox{10cm}{!}{\includegraphics{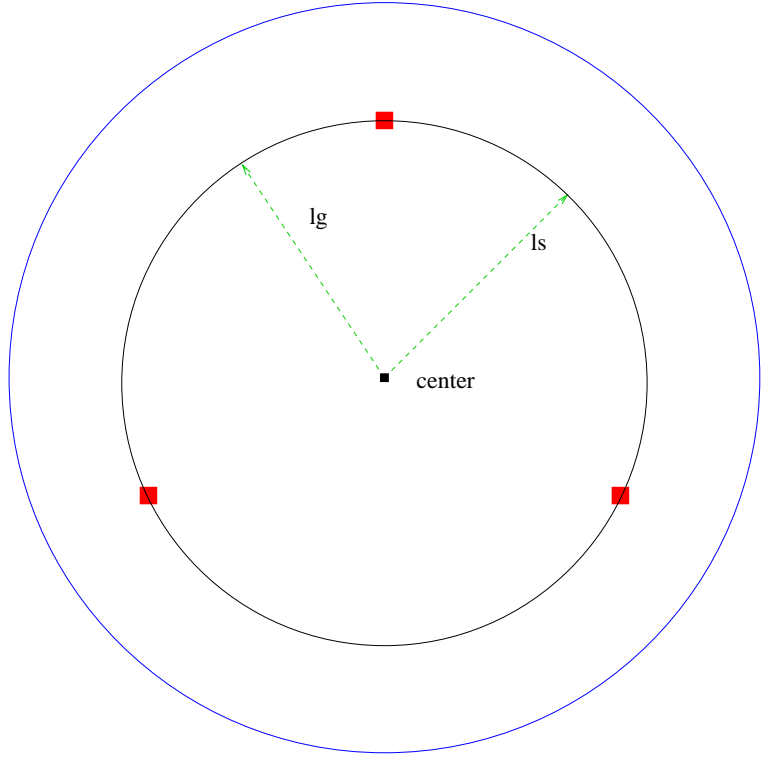}}
  \caption{Our acute triangle, in the $t=0$ plane.  The causal set is sprinkled into the
 causal interval with endpoints $(t=\pm 6.5614, x=y=0)$.  The
    large blue circle is the intersection of the two lightcones which form
    the interval, and hence has radius 6.5614.}
  \label{acute_triangle}
\end{figure}

\begin{figure}[hbtp]
  \psfrag{deviation}{$\left<\frac{l - l_c}{l_c}\right>$}
  \psfrag{lg deviation}{\hspace{-21.5mm}$\left<(l_g - 4.9193)/4.9193\right>$}
  \psfrag{ls deviation}{\hspace{-20mm}$\left<(l_s - 4.9193)/4.9193\right>$}
  \psfrag{log2 N}{$\log_2 N$}
  \psfrag{0}{\hspace{-5mm}0}
  \includegraphics{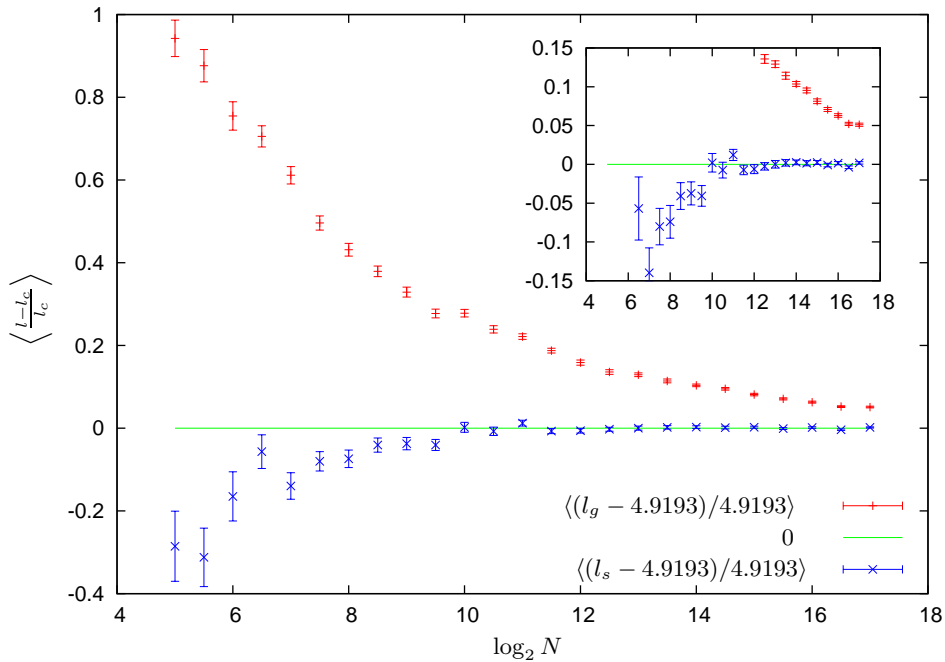}
  \caption{Performance of diameter measures for an `acute triangle', on
    causal sets sprinkled into an interval of $\M^3$.  On the vertical axis,
    $l$ stands for $l_g$ or $l_s$, while $l_c$ stands for 4.9193,
    respectively.  The inset plot zooms in on the asymptotic behavior.}
  \label{acute_triangle-voldist}
\end{figure}

Figure \ref{acute_triangle} illustrates our acute triangle. For this
configuration, the continuum values for the circumscribing and
bounding sphere coincide at $l_g = l_s = 4.9193$. In
figure \ref{acute_triangle-voldist}
we demonstrate the following results: Both
$l_g$ and $l_s$ converge to the continuum value 4.9193 (and therefore in
the limit they coincide), but we can see from the inset
that $l_s$ converges more quickly
than $l_g$.

\subsection{Remarks}

Let us summarize and stress here a few issues before we move to a (quite)
different proposal.

\begin{itemize}
\item[(a)] The continuum distances $l_g$ and $l_s$ coincide if all the `included
angles' at the spacelike elements are `acute' (or generally when the
circumscribing and bounding sphere coincide).
\item[(b)] The accuracy of $l_s$ does \textbf{not} depend on the included angles
at the elements, and continues to give accurate results for
obtuse angles.
\item[(c)] Even when $l_s$ and $l_g$ agree in principle (for acute
angles) $l_s$ gives a (slightly) more accurate value.
\item[(d)] By comparing $l_s$ and $l_g$ we can deduce whether, for example, the
triangle (in 2+1-dimensions) formed by the elements is an obtuse or
acute triangle. We therefore have, for acute triangles $l_s=l_g$,
while for obtuse, $l_s>l_g$.
Furthermore, one can deduce at which vertex the obtuse angle occurs, by
examining the timelike distances $d(w,x_i)$ and $d(z,x_i)$ involved in the
$l_g$ computation.  The element $x_i$ which is furthest from $w$ and $z$ will
be the one with the obtuse angle.
\item[(e)] $l_g$ fails for obtuse triangles, in a similar manner as \bgdist,
  but much less severely.  In particular, it does not converge to zero and in
  practice we expect the error to be usually quite small.
\item[(f)] $l_s$ cannot be generalized easily in curved spacetimes,
  because it relies on (\ref{l_s function of l_m}), which was derived in flat
  spacetime. It cannot even be used for detecting nearest neighbors in a
  curved spacetime, despite the fact that for closest neighbors we can expect
  that things would be approximately flat.  This is
  because we use equidistant elements, which may be far away, even for
  ``nearby'' unrelated elements.
\end{itemize}

We also mention three possible uses of the distances $l_g, l_s$ in order to
recover more 
geometric structure from the causal set.

\begin{itemize}

\item[(i)] We can attempt to define \emph{proximity} relations.
For example, we can define balls around each element, 
and from this we may recover some of the underlying topology.

\item[(ii)] Given a particular 
  inextendible antichain $A$ (the discrete analogue of a spacelike
  hypersurface), one could perform the following procedure in order to find
  the element closest to any other given element.
  Select an arbitrary element $x \in A$.  Find the element
  $y \in A$ such that there does not exist any element $z \in A$ that makes
  $xyz$ an obtuse triangle. This element $y$ is then the closest neighbor to
  $x$ within $A$ \cite{Brightwell}.

One could repeat this procedure $d$ times (for an antichain derived from a
causet faithfully embeddable into $\M^d$) for each element $x \in A$, to form
a $d$-valent adjacency graph for $A$.  This graph then could be used for
various purposes, \eg\ to define spatial distance as the length of the
shortest path connecting a pair of elements $x$ and $y$.

\item[(iii)] Given a pair of unrelated elements $x,y$, one may attempt to
  define a spacelike distance in $\M^3$, by considering any element $z$ that
  makes $xzy$ an obtuse triangle with the obtuse angle at $z$. In that case
  $l_g$ simply gives the `longer leg' of the triangle, which is the distance
  between $x$ and $y$. However this suggestion may fail because it depends on
  the third element. Minimizing over all possible third elements may bring
  back a similar problem as occurs with \bgdist{}.  Even if one selects a
  third element randomly (such that an obtuse angle occurs at $z$), or
  averages over all possible such elements $z$, there is
  still the underestimation as mentioned above.\footnote{Note that these
    schemes would require identifying all of the infinite number of elements
    that form an obtuse angle.  In addition, in higher dimensions, one would
    have to compose the appropriate generalization of this construction.
    These reasons make it not as appealing as the approach detailed in
    section \ref{Section 5}.}
\end{itemize}

\section{Spatial distance} \label{Section 5}

After having explored the difficulties with previous approaches, and obtained
certain geometric information with some new constructions, we proceed by
considering a quite different proposal, which is the main result of the paper.

\subsection{Motivation and Definition}

We come back to the initial problem of defining spatial distance. As
explained in section \ref{Section 2}, it stems from the
existence of many (in fact infinite) ``minimizing pairs''. In the
previous section, we explored the possibility of obtaining some
geometrical information by considering multiple ($d$) elements, whose
common lightcones would intersect at a unique point. Here we take a
different stance initiated by the following important observation.
\emph{Each minimizing pair (which lies close to a continuum minimizing pair)
gives on average the correct expected distance.}
All the problems arise when we minimize over these pairs. To use this
observation to define a distance, we will need the following:

\begin{itemize}

\item[(a)] a mechanism to select causet elements which lie close to a
  continuum minimizing pair.  Such a mechanism involves
\begin{itemize}
\item[(i)] finding elements which are close to the intersection of the future
  (or past) light cones of our unrelated pair of elements
\item[(ii)] for each such element $z$ in the future (say), select an element
  $w$ in the common past which locates a pair $(w,z)$ which is close to some
  continuum minimizing pair.
\end{itemize}

\item[(b)] 
to take an \textbf{average} over minimizing pairs, and not minimize

\end{itemize}

As already described in section \ref{Section 4}, 2-links will locate causet elements
close to the intersection of the pair of light cones.
Naively one may 
ask that, given a future 2-link $f$, select the past 2-link
$p$ that minimizes the timelike distance $d(p,f)$. However, this is unlikely to
give a correct answer.
Recall that in the continuum, for a given point on
the common future lightcone, there exists a unique point on the common past
lightcone that minimizes this distance.
Future and past 2-links lie approximately on
the intersection of the lightcones, but they are rare.
Given a future 2-link $f$,
it seems clear
that in general there will be no past 2-link
near the point on the past common lightcone that would minimize
the distance $d(p,f)$.   Therefore we propose the
following procedure: 

\begin{itemize}
\item[Step 1:] Given spacelike elements $x,y$ we find a future 2-link $f_i$.
\item[Step 2:] Find the element $p_i$ in the common past of $x$
and $y$ that makes the timelike distance $d(p_i,f_i)$ minimum.
This will select a minimizing pair.
\item[Step 3:] Store the timelike distance $d^i(x,y)$ for the future 2-link $f_i$.
Note that this distance can be calculated either as
the length of the longest chain, or by using \voldist.
\item[Step 4:] Repeat for all other future 2-links. 
\item[Step 5:] Take the average over all future 2-links $\langle d^i(x,y) \rangle$ to be the spacelike
  distance between elements $x$ and $y$.  We call this average the
  \emph{2-link distance} between $x$ and $y$.
\end{itemize}

To give some feel for the importance of using 2-links here, consider what
would happen if in Step 1 we find a minimal element $z$ to the common future
of $x$ and $y$, instead of a future 2-link.  This element would be close to
one of the future light cones of $x$ or $y$, but not the other.  Step 2 will
yield a symmetrically placed element $w$ in the common past.  It is not
difficult to see that the proper time separation between this pair $w,z$ can
be made arbitrarily large, by selecting an element $z$ as far as one likes
from the intersection of $x$ and $y$'s future light cones.  Thus the 2-link
construction plays a crucial role here in recovering the correct spatial
distance.

For infinite Minkowski space, we expect to find an infinite number of 2-links,
and thus this
corresponds 
to taking an average over an
infinite collection of (almost) independent random variables, each of which comes from the same
underlying distribution. This gives a
better and better approximation as we consider more pairs, and should 
give a well defined value in general (by the \emph{central limit theorem} and
the \emph{law of large numbers}).
To be complete,
we must specify the order in which we add the terms in the average, in order
for the infinite sum 
to be well defined.
In our simulations (section \ref{2-link_numerics}), we
consider future 2-links in order of their embedded time coordinate.  Since the
sprinkling domain is finite, it of course makes no difference.  It seems clear
that any `reasonable' ordering of 2-links in this sum will work.  We
conjecture in particular that the order we use in the simulations will give a
mean which converges to the spatial distance in the embedding, for a faithful
embedding into infinite Minskowski space.  
More precisely, given a
causal set which faithfully embeds into Minkowski space, with embedding
$\phi$, and any unrelated pair $x$ and $y$, first select an arbitrary frame in
which $\phi(x)$ and $\phi(y)$ are simultaneous, and then add the terms
$d^i(x,y)$ above in order of increasing $t$-coordinate of $\phi(f_i)$.
Alternatively, one can define an ordering of the terms
that is independent of any embedding.
For a finite region on Minkowski space, these subtleties
do not arise.

Note that we could have used past 2-links instead and minimized
over the common future. For a more symmetrical definition, we could
average over both. With this procedure we evade the problem
mentioned above, that in general there will be no past 2-link exactly on the
point of the common past lightcones that would minimize the distance.
Instead of looking to past 2-links, we simply select an element
close to the relevant continuum point, by minimizing the geodesic distance. We
are guaranteed to get a
non-trivial result because, given a future 2-link, there exists a
unique point in the continuum, to the common past, that minimizes
the timelike geodesic distance.
We avoid getting a degenerate distance
in a similar way that we do not have problems in
$1+1$ dimensions where there is a unique continuum pair.
Finally, taking average over all the 2-links makes the distance
less subject to fluctuations.

An important note is that for this procedure to work we need the 2-links to
exist. As we have pointed out earlier (section \ref{Section 4}), this is
indeed the case for all dimensions greater than $1+1$. We could attempt to
modify the definition in order to incorporate the $1+1$ case as well, but this
may not be 
important, since in $1+1$ the \bgdist{} works anyway\footnote{The modification
  would be to consider ``closest 2-neighbors'', (defined suitably), rather
  than 2-links. For the case that there exist 2-links, the definition of
  ``closest 2-neighbors'' would coincide with 2-links.}.

Before proceeding to see how this works with some simulations, we should
point out that this distance function will overestimate the actual
distance. 
This is the same effect that
we called \spover.  This overestimation should always be comparable to the
`discreteness scale' (the spacing between embedded points), and thus should be
negligible at large distances.
In addition, the distance between two given elements should
not be affected by considering portions of the causal set (faithfully
embeddable into Minskowski space) which are far from the elements (in some
given frame), unlike with \bgdist{}.
Therefore, `without loss of generality', we can consider a finite section of the full causal set,
and get answers that should not be different from infinite causal sets.

\subsection{Numerical results and comparison with \bgdist{}}
\label{2-link_numerics}

The following plots compare 2-link distance with \bgdist.

\begin{figure}[hbtp]
  \psfrag{2-link distance}{\hspace{-3mm}2-link distance}
  \psfrag{0}{0}
  \psfrag{bgdist}{\hspace{-24mm}\bgdist}
  \psfrag{deviation}{$\left<\frac{X - X_c}{X_c}\right>$}
  \psfrag{log2 <N>}{$\log_2 \la N \ra$}
  \includegraphics{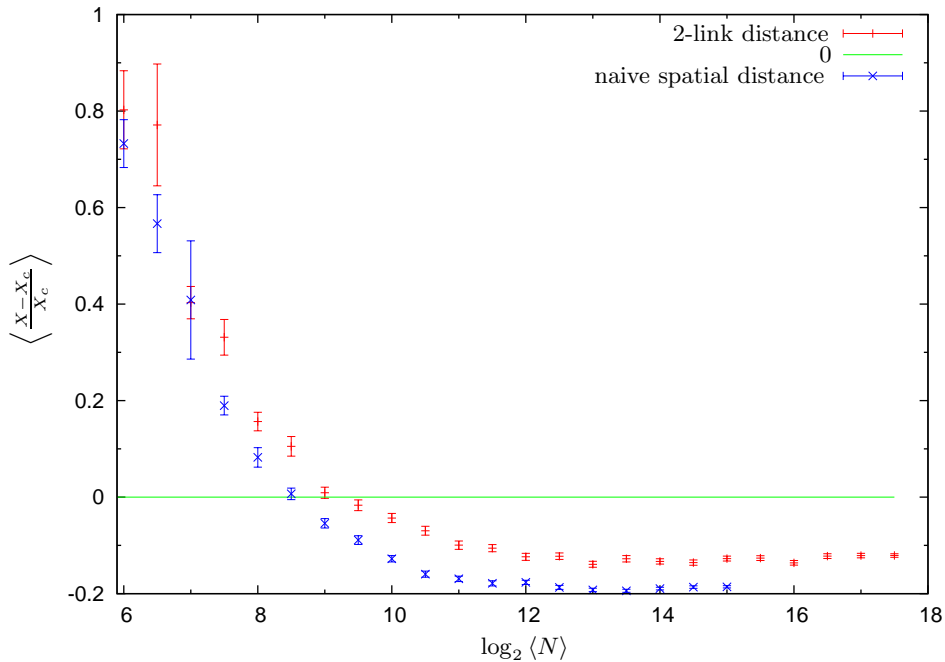}
  \caption{Comparison of 2-link distance with \bgdist, for sprinklings into a
    fixed cube $[-1,1]^3$ in $\M^3$, for various densities $\rho \to \infty$.
    Target points are fixed at $(t=0, x=\pm \frac{3}{10}, y=0)$.}
  \label{2linkdist}
\end{figure}

The results shown in figure \ref{2linkdist} are for simulations analogous to
those whose results are shown in figure \ref{bgdist3d}.
We send the
sprinkling density to infinity, while measuring the distance between
two elements which are sprinkled near fixed target points 
at $t=0, x=\pm 3/10, y=0$, as 
described in section \ref{bgdist_sec}.
Using coordinates for which the sprinkling density $\rho$ is fixed at 1,
this 
means that we increase the distance between the target points, 
and 
observe that the 2-link distance
converges to some constant small error, as the
distance goes to infinity. For comparison we show the results of \bgdist{} on
the same plot.
We see that it behaves similarly.
The effect of \spover{} decreases rapidly for larger causal sets
(larger distances in fundamental units). The \bgdist{} is always slightly less than the 2-link distance,
since by definition it is obtained by
taking the minimum over pairs of elements in the common past and future, so it
cannot be less than a distance measure which employs an average over some
subset of these pairs.
The underestimation of \bgdist{}, that leads to its
failure, is due to the existence of many (infinite) ``minimizing
pairs''. However, given the way we select the sprinkling region, we take into
account only relatively few independent minimizing pairs, and increasing $N$
does not add more pairs because, in fundamental units,
the distance also increases.
Therefore, we do not expect
to see an increasing discrepancy in the results.
Note that both
distances underestimate. This is likely due to \tiunder{}. 
Here we have used an asymptotic value for $m_3 = 2.27845$.

\begin{figure}[hbtp]
  \psfrag{2-link distance}{\hspace{-5mm}2-link distance}
  \psfrag{0}{0}
  \psfrag{bgdist}{\hspace{-24mm}\bgdist}
  \psfrag{deviation}{$\left<\frac{X - X_c}{X_c}\right>$}
  \psfrag{log2 T}{$\log_2 T$}
  \includegraphics{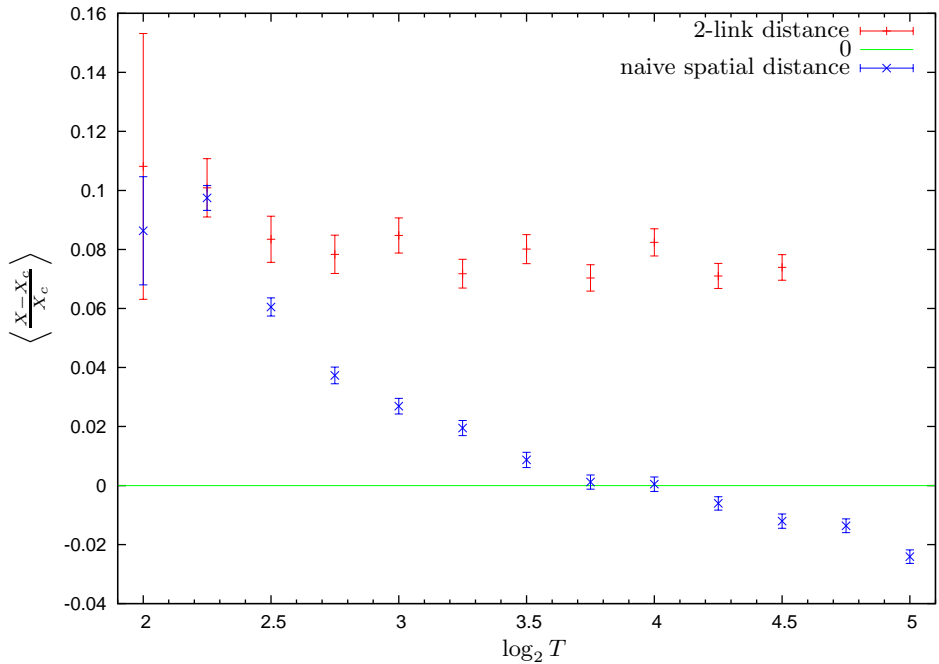}
  \caption{Comparison of 2-link distance with \bgdist, for causets sprinkled
    into an expanding region $t \in (-T,T), x \in (-4,4), y \in (-T,T)$ in
    $\M^3$, with target points at $(t=0, x=\pm 4, y=0)$.  In these coordinates
    $\rho$ is fixed at 1.}
  \label{bgfailure-2linkdist}
\end{figure}

In figure \ref{bgfailure-2linkdist}, we compare the performance of 2-link
distance with \bgdist{}, for the same sorts of causal sets used to generate figure \ref{bgfailure}.
The two elements (or target points) are separated by a constant distance (in
fundamental units), while we increase the volume of the region of Minkowski space
into which we sprinkle. 
This results in considering more and more (independent) minimizing pairs. Here again 
we see that 2-link distance is not at all affected by the
increasing size of the sprinkling region, 
as expected. The 
\bgdist{}, however, decreases as the sprinkling region grows, as noted earlier.
This is the effect of taking into
account more and more independent minimizing pairs, and selecting always the
pair that gives the smallest distance. A good distance function
should not be affected by the size of the region of Minkowski space into which
we sprinkle (at least beyond some reasonable lower bound), 
when the two spacelike target points remain unchanged.

A final note on figure \ref{bgfailure-2linkdist} is that the 2-link
distance seems to overestimate the continuum distance. This is
expected, since in fundamental units all the points of this plot
correspond to a very small distance ($\ltsim 8$) (we can also compare with figure
\ref{2linkdist} the first points in the plot). For computational
capacity reasons we could not do the same (increasing size of the sprinkling
region, while holding the target points fixed) for a much larger distance (in fundamental units).

\section{Toward curved spacetime 
}\label{Section 6}

\subsection{Nearest neighbors}

Having obtained a spacelike distance for causal sets which faithfully embed into Minkowski space, we turn to
the question of whether we can define closest spatial neighbors. In the
timelike direction, the closest neighbors are links. In a (somewhat) analogous
manner we regard the spacelike neighbors of an element $x$ as those whose 2-link
distance from $x$ is less than a small threshold.
More formally, we wish to define a second, symmetric `\slink{}' relation on
the elements of the causal set.
\newline\textbf{Definition 9:}
An \emph{\slink{}} is the relation between a pair of
elements of a causal set whose 2-link distance is less than a fixed threshold
$\threshold$.\\
Any choice of $\threshold$ (and whether to use \voldist{} or the length of the
longest chain to measure timelike distance), will define the `\slink{} relation' on
a causal set.
By definition, the minimum possible value for 2-link distance is 2.
Even for neighboring elements in the embedding, it is extremely unlikely that
their 2-link distance will be exactly 2, as the fraction of 2-links which
contribute anything to the sum in the mean besides 2 would have to be
vanishingly small in the limit of infinite Minkowski space.
We seek more `typical' neighbors, and thus choose a threshold by the following
procedure.\footnote{Note that this procedure is not necessary, any reasonable
  value of $\threshold$ (say below 3) can be used in practice.}
\begin{enumerate}
\item Sprinkle $N$ elements into a cube $[-1,1]^3$ in $\M^3$ with $\langle N \rangle$ = 5793.
\item Use the origin as a target point to select an element $x$.
\item Use the embedded location of this element as the target point to select
  a second element $y$.  Here we take the closest $y$ which is unrelated to $x$.
\item Repeat for 400 sprinklings, and compute the mean 2-link distance $\mu$
  and its error $\delta\mu$.  
\item 
Set $\threshold = \mu + \delta\mu = 2.186178$.
\end{enumerate}

\begin{figure}[hbtp]
\scalebox{2.10}{
  \psfrag{ 0.8}{}
  \psfrag{ 0.6}{}
  \psfrag{ 0.4}{}
  \psfrag{ 0.2}{}
  \psfrag{ 0}{}
  \psfrag{-0.2}{}
  \psfrag{-0.4}{}
  \psfrag{-0.6}{}
  \psfrag{-0.8}{}
  \includegraphics{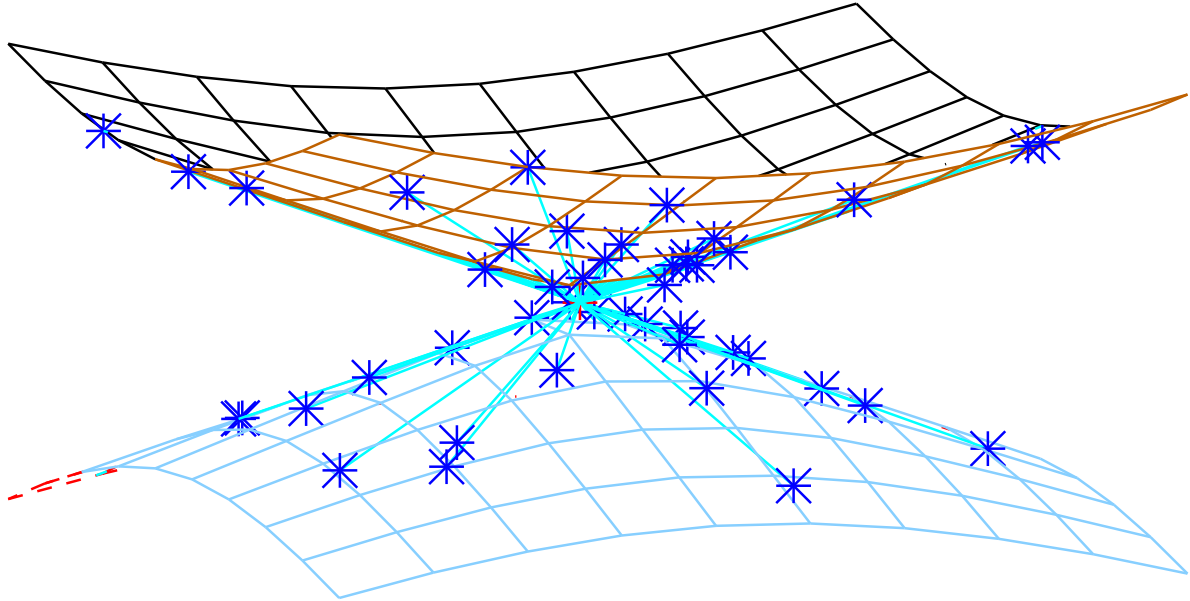}}
  \caption{Spatial nearest neighbors of an element in a sprinkling into (a
    fixed cube in) $\M^3$.  $\langle N \rangle = 65\,536$.  The future and
    past light cones of the `origin element' $x$ are shown.  The spacelike
    cyan lines are drawn between $x$ and each neighbor, for emphasis.}
  \label{neighbors}
\end{figure}

Since we already have a notion of closest neighbor for related
elements (namely links), we now have a definition for closest neighbors in
general.
Figure \ref{neighbors} shows how the \slink{}s are
distributed for a sprinkling into a cube of $\M^3$ with $<N> =
65 536$. The element closest to the center of the cube is our reference, and the
blue elements are all the \slink{}s from that element. As expected, they
tend to lie on a hyperboloid.

\subsection{Spatial distance in curved spacetime}
\label{curved_spacetime_sec}
In the continuum, one defines a curve as a smooth map from a
connected subset 
of the real line 
to the manifold $\mathcal{M}$.
The analogue of a curve, in a causal set, would be a sequence of
elements, \ie\ a map
from (a subset of) $\textbf{N}$ to the causal set $\causet$.

It is obvious that the smoothness condition cannot be implemented.
However, we \emph{can} have a continuity condition. Intuitively, we
require that the curve be composed of a sequence of elements for which each
successive pair are `nearest neighbors', in either the timelike (link) or
spacelike (\slink{}) sense.
To properly define continuity, one needs to specify
the topology 
on the causal set
$\causet$, and on the natural numbers $\textbf{N}$.
However, an obvious problem is that both domain and range are ``locally
finite'', which implies that natural choices (\eg\ order topology or open
ball topology respectively) will lead to the discrete topology in which case
\emph{any} map is continuous. The correct construction of the appropriate
topology, which despite being discrete captures continuum notions, is
described in \cite{Sorkin:1991a}, and similar arguments were also used in
\cite{homology}. The result of that construction applied to
our case is that a curve is ``continuous'' if it maps any two consecutive
numbers of the $\textbf{N}$ to elements in $\causet$ that are related as
``closest-neighbors'', \ie\ are either links or \slink{}s. Further details on
this are beyond the scope of this paper and the reader is referred to the
original references.

Having defined what we mean by continuous curve, we can easily
define the length of a curve, as simply its cardinality. An important thing to
note here is that we have now
defined the length of continuous curves in a causal set, which
need \emph{not} be solely timelike
or solely spacelike, but can also be mixed, \ie\ at some points spacelike and
other points timelike.\footnote{In \cite{homology}
the spatial topology of a causal set was recovered (with drastically different
methods than ours). 
The possibility of defining
continuous spacelike curves,
something that was not possible earlier, could be recovered.  However the case
of general continuous 
curves is something that could not be obtained.}

So far we have only considered causal sets that are well approximated
by Minkowski spacetime. One of the virtues of this discussion is
that it could be generalized for curved spacetimes. In particular,
assuming that the spacetime is approximately flat at some small scale above
the Planck scale but much below the
macroscopic scales (\ie\ locally
flat), we can use the above construction to define
closest neighbors (\slink{}s).  The idea would be to consider some region,
for example an order interval, whose size is comparable to this local
flatness scale.  One would then sum over only the 2-links which lie within this
local region.
Since the determination of an \slink{} in this manner depends only on a local subset of the
causal set (in any frame for which local flatness is valid), it seems likely that
any construction that is based on
the closest neighbors, such as length of a continuous curve, will carry over to
(causets which faithfully embed into) curved spacetime.
A very similar strategy has been employed to write down an action functional for a
causal set with a scalar field \cite{sverdlov}.  
See also \cite{Gibbons}, for a similar approach to extracting geometry from
the causal structure of a
locally flat region.
Note 
that in curved spacetime
the concept of spacelike distance between two points is not 
well defined in general, but the length of a (smooth) curve is.

If we restrict attention to an inextendible antichain, analogous to a spatial
hypersurface,
the definition of closest
neighbors can give a spatial distance, defined as the graph
distance between two elements, where the graph edges coincide with the \slink{}s
between elements of the antichain.  This distance is 
what one would expect
from minimizing the length of the possible (continuous) 
spacelike curves. 
Note that there is a direct analogy with the continuum case (where spatial
distance between two points is indeed the length of the minimum curve between
those points).

\begin{figure}[htbp]
  \includegraphics[width=\textwidth]{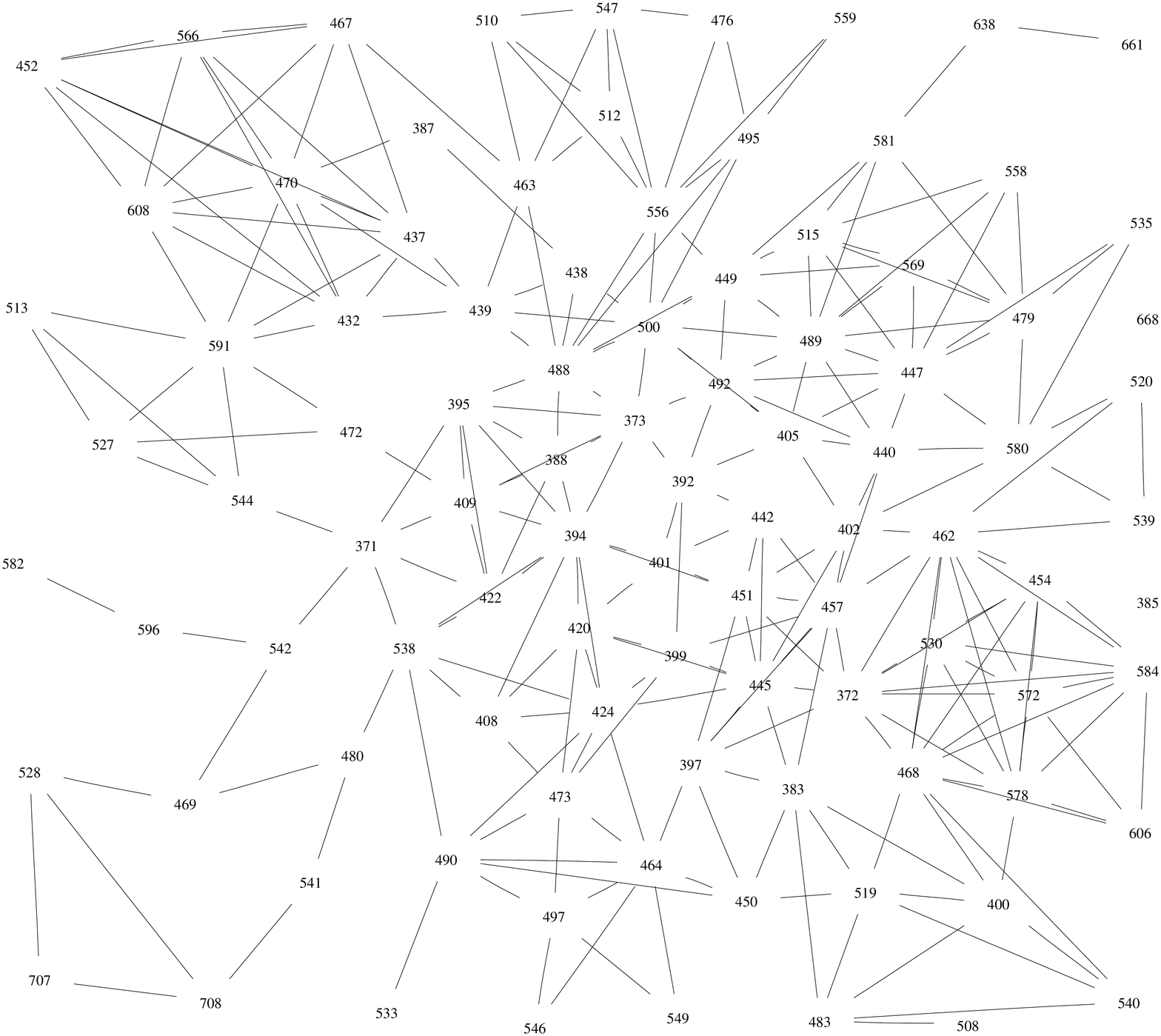}
  \caption{Graph formed from nearest neighbors in a `smooth' inextendible
    antichain, at the center of a causal set faithfully embedded into a box in
    $\M^3$, of spacetime volume 1024.  The nearest neighbor threshold is set
    to 2.7.}
  \label{spatial_slice}
\end{figure}

Figure \ref{spatial_slice} depicts such a graph, for a
`smooth' inextendible antichain at the `center' of a $\la N \ra = 1024$
element causal set faithfully embedded into a cube in $\M^3$.\footnote{The
  inextendible antichain was generated using the thickening technique
  described in \cite{tas}.  Starting from the minimal elements, we
  thicken to $v=98$, and take the maximal elements $A$ of this thickened
  antichain.  Since in general this will not form an inextendible antichain,
  we must extend it by adjoining the minimal elements of the complement of $A
  \cup \fut(A) \cup \past(A)$.}  Since the inextendible antichain has much
fewer elements than the causal set, it is necessary to use a larger threshold
to define adjacency, as otherwise very few pairs will be regarded as adjacent
in the antichain.  Here we use 2.7.  The picture of the graph was generated by
the Graphviz package \cite{graphviz}.

Finally, one can speculate wildly, about how one may be able to
recover tangent spaces and thus the full  metric on a causal set.
This is left for future work, but here we just point out some
observations and a potential problem.

One could consider that for a given element $x$, each of the closest neighbors
are possible ``vectors'', since we can consider them as equivalence
classes of continuous curves (each with the same tangent at $x$).
Moreover, if we 
consider the limit of infinite sprinkling density,
we expect that these vectors will form a dense set. 
Therefore in some sense we cover all the directions, at least in the large
density limit.
However, for all this to make 
sense, we need to define the
tangent space as a vector space 
(in order to define
vector fields, and thus a metric tensor).  To do this one must somehow define
a closed operation of vector addition, without resorting to the background structure. 
Perhaps one could make progress
by considering the angles between vectors (which could be
done using our definitions of 
timelike and spacelike distances).

\section{Summary and conclusions}\label{Section 7}

We have discussed various ways of recovering geometrical
information of continuum spacetime from a causal set that the
spacetime approximates. All of the constructions involved
causal sets
that are approximated by Minkowski spacetime, however some remarks about
generalization to curved spacetime appear in section \ref{curved_spacetime_sec}.

In particular we began by reviewing and numerically testing some
previous definitions (\eg\ from \cite{bg}), namely on timelike and spacelike
distance.  
The timelike distance between two elements is
proportional to the length of the longest chain connecting them (for large $N$).
From our simulations we extracted the asymptotic value of the
proportionality constant $m_3$ ($m_2$ is known to be 2)
and computed the value of $\meff{d}$ (for
$d=2,3$), which is an effective proportionality constant
relevant for finite timelike
distances. Furthermore we examined the \bgdist{}, which is defined
to be the smallest timelike distance between the common past and
common future of the two unrelated elements in question. As argued in
\cite{bg}, this fails to
recover the continuum spacelike distance, due to the
existence of infinite ``independent minimizing pairs''.  We observed
evidence of this failure numerically for finite causal sets.

The second part of our paper concerned some new ideas. First we
defined the concept of an $n$-link. This is defined to be any element
that is linked to $n$-unrelated elements. It essentially captures the
notion of points lying on the light-cones of \emph{all}
$n$-unrelated (\ie\ spacelike) elements. It was shown that in $d$
dimensions, for $n$ unrelated elements, where $n\leq d-1$, we find an
infinite number of $n$-links, while for $n\geq d$ we generally find none (for
a given $n$ elements). This can help us determine the dimension of the
causal set, since the smallest $n$ for which we find no $n$-links is
equal to the dimension ($n=d$).
It may also serve as an indicator of `manifoldlike' causal sets, as the
numbers of $n$-links should behave in this particular manner for faithfully
embeddable causets.

For $d$ unrelated elements in $\M^d$
we defined the ``\lgdist'' $l_g$, which corresponds to
the diameter of the bounding sphere of those elements.
Using the
concept of \emph{equidistant} elements, we defined the ``sphere
distance'' $l_s$, which computes the diameter of the \emph{circumscribing}
sphere of the $d$-unrelated elements. In 2+1 dimensions, noting that the circumscribing
and the bounding spheres coincide when all the angles at the
elements are acute, we can deduce
whether there is an obtuse angle by comparing $l_g$ with $l_s$.
In arbitrary dimension $d$, the statement that all angles are acute can be
expressed more generally as the statement that the center of the $d-2$-sphere
defined by the $d$ elements 
lies within their convex hull.

The third part of this paper (and the main result), was to define
```spatial distance'', and furthermore to define spatial closest
neighbors, called \emph{\slink{}s}. This was done by making the
observation that each minimizing pair gave the correct distance, and the
problem arose when we minimized over all those pairs. In this
proposal, we had to (a) select relevant pairs, and (b) take an average over
their (timelike) distance.
To select the relevant pairs we used crucially the concept of a $2$-link, that
lies close to the intersection of the lightcones.
We then numerically tested this suggestion in $\M^3$, and found it to
give good distances (close to the continuum ones), and most importantly it
did \emph{not} suffer from the same problem of underestimation of
the \bgdist{} (\cf{} figure \ref{bgfailure-2linkdist}). We then used
this definition to define ``\slink{}s'', being the analogues of
links for unrelated elements. In finding the \slink{}s, we are working
locally, and thus we may expect that these considerations can be
generalized directly to curved spacetimes. In both cases
(flat/curved) knowing the closest neighbors (timelike and spacelike)
allows us to define the length of a (continuous) curve. This is an important
step to the recovery of the full curved spacetime geometry.

Given such a full recovery of spacetime geometry from a causal set (which is
yet to be achieved), one could likely do the following.
\begin{itemize}

\item[(a)] Prove the ``Hauptvermutung'', which is the main
conjecture of the causal set program. It claims that two distinct,
non-isometric spacetimes cannot arise from a single causal set.

\item[(b)] Given a causal set, we will be in position to tell if it
corresponds to a spacetime, and which is the corresponding metric.
This will guide us as to whether the quantum dynamics we impose
leads to physical predictions that agree with observation!

\item[(c)] It would be possible to write down the Einstein action in
terms of causal set quantities. This 
may be used to
``naively quantize'' by computing sums 
over causal sets with the appropriate amplitudes.

\end{itemize}

On a more immediate time frame,
these extra structures that we can compute from 
a causal set could be useful in a number of ways. One could, for example, use
these (\eg\ the $n$-links, spatial distance, or the length of curves) to more
readily derive entropy bounds and compute black hole entropy from causal sets
(as in \cite{entropy-causet}), because they allow one to more readily identify
the relevant substructures of the causal set to count.

\section{Acknowledgments}
We are especially grateful to Graham Brightwell for a number of
discussions on this work, for example in leading us toward the relevant
parameters required to demonstrate the failure of \bgdist{} on the computer.
We are also grateful to Rafael Sorkin, David Meyer, and the attenders of
`relativity lunch' at Imperial College London, for numerous helpful discussions.
We thank the anonymous referees for numerous constructive remarks and
suggestions.

We thank Yaakoub El-Khamra for his patient assistance with our parallel Monte
Carlo code.

This research was supported by a number of grants/organizations, including
the Marie Curie Research and Training Network ENRAGE
(MRTN-CT-2004-005616), the
Royal Society International Joint
Project 2006-R2,
and the Perimeter Institute for Theoretical Physics.
Research at Perimeter Institute is supported by the Government of Canada
through Industry Canada and by the Province of Ontario through the Ministry of
Research \& Innovation.

Many of the numerical results were made possible by the facilities of the
Shared Hierarchical Academic Research Computing Network
(SHARCNET:www.sharcnet.ca).

PW thanks the Perimeter Institute for Theoretical Physics for hospitality.

\appendix
\section{Proof of existence of equidistant elements in 1+1 dimensions}
\label{proof of equidistant points}

Here we prove the existence of equidistant elements in 1+1
dimensions. In higher dimensions the proof gets much more complicated.  We
rely on 
numerical evidence and the intuition obtained from the 1+1
dimensional case to suggest that they also 
exist generically in higher dimensions.

We are looking for equidistant elements in 1+1 dimensions between causet
elements mapped to points  $x_1=(-r,-r)$ and
$x_2=(-r,r)$, for some faithful embedding. First we note is that in 1+1
dimensions the order intervals between a point $y$ in the common future
($J^+_c:=J^+(x_1)\cap J^+(x_2)$) and 
$x_1$ and $x_2$ are divided into one common region and two disjoint
regions. For an element to be equidistant, 
since there are obviously an equal number of elements in the common region,
it suffices to show that the independent regions
have same number of elements mapped to them.  These independent regions are
defined by the
coordinates of the point $y$ in the common future.

Let us move to lightcone coordinates for which $x_1=(-\a,o)$, $x_2=(0,-\a)$
and $\a=r\sqrt2$. Given a point in the common future with coordinates
$y=(u,v)$, it is clear that the area of the region
$[x_1,y]_c\setminus[x_2,y]_c$ (\ie\ the area causally between $x_1$ and $y$
after we subtract the overlap with the region $[x_2,y]_c$) is equal to $V_1=\a
u$ and similarly $[x_2,y]_c\setminus[x_1,y]_c$ has area $V_2=\a v$. Note again
that the contents of $[x_1,y]_c$ and $[x_2,y]_c$ are independent for any $u$
and $v$. We can now compute the following:

\begin{itemize}
\item[(a)] The expected number of elements $E(n\in[x_1,y]_c)$ in the common future
  $J^+_c$, such that the region $[x_1,y]$
  contains exactly $n$ elements\footnote{using that
    $\int_0^\infty x^n\exp{(-ax)}dx=\frac{n!}{a^{n+1}}$.} :

\be
\langle E(n\in[x_1,y]_c)\rangle =
\int_0^\infty du dv \frac{(\a v)^n}{n!}\exp{(-\a v)}=\int_0^\infty du
\frac1\a\rightarrow\infty
\ee

\item[(b)] The expected number of elements in the common future $J^+_c$, such
  that the region $[x_1,y]$ contains exactly $n$ elements \textbf{and} the
  region $[x_2,y]$ 
  also has exactly $n$ elements:

\begin{eqnarray}
\langle E(n\in [x_1,y]_c \textrm{ and } n\in [x_2,y]_c) \rangle &=&
\int_0^\infty dv dv\left(\frac{(\a v)^n}{n!}\exp{(-\a
v)}\right)\left(\frac{(\a u)^n}{n!}\exp{(-\a
u)}\right)\nonumber\\ &=& \frac1{\a^2}
\label{equidistance_n}
\end{eqnarray}

\end{itemize}
The expected number of equidistant elements $E(\textrm{equidistant})$ is then
recovered when we add (\ref{equidistance_n}) for all possible values of $n$, in other words
\beq
\langle E(\textrm{equidistant})\rangle=\sum_{n=0}^\infty
\frac1{\a^2}\rightarrow \infty \;.
\label{equidistance_sum}
\eeq
This completes the proof of the
existence of infinite equidistant elements in 1+1 dimensions.

\begin{itemize}
\item[Note 1:] In higher dimensions the proof is more difficult
since the simplifying features of using the lightcone coordinates do
not apply.
\item[Note 2:] $\frac1{\a^2}\propto \frac1r$, so if $r\rightarrow\infty$
(\ref{equidistance_sum}) is not well defined. For any finite distance between
the two points, the above argument holds.
\item[Note 3:] Both points (a) and (b) above are to be expected.
Intuitively there will be some $v$ such that $[x_1,y]_c$ contains
exactly $n$ elements.  On average we expect that $v\a=n$.
One can then find another $y'$ whose $v$ coordinate is $v + \delta v$, such that $[x_1,y']_c$ contains
exactly $n+1$ elements, and on average $(v+\delta v)\a=n+1$.
Thus we expect on average $\delta v=1/\a$.
However any element in the whole region between $v$
and $v+\delta v$ for \textbf{any} $u$ will give us elements which find
exactly $n$ elements in $[x_1,y]_c$. Since the above area is infinite, the
expected number is infinite.

An analogous argument gives an expected value for $\delta u=1/\a$.
If we require that both $[x_1,y]_c$ and $[x_2,y]_c$
have exactly $n$ elements, we end up considering the area $\delta u
\cdot\delta v=\frac1{\a^2}$. Thus the expected elements in $J^+_c$ that
have exactly $n$ elements in both $[x_1,y]_c$ and $[x_2,y]_c$ comes naturally to
be $1/\a^2$.

\end{itemize}

\bibliography{apssamp}

\end{document}